\documentclass[useamsfonts]{pasj01}
\usepackage{amssymb}
\usepackage{natbib}
\usepackage{footnote}
\usepackage{cleveref}
\usepackage{threeparttable}
\begin{document}

\title{Investigation of the Origin of the Anomalous Microwave Emission in Lambda~Orionis}

\author{Aaron C. Bell\altaffilmark{1,2}}
\altaffiltext{1}{University of Tokyo, Graduate School of Science, 7-3-1 Hongo, Bunkyo-ku, Tokyo, Japan 113-0033}
\altaffiltext{2}{Ridge-i Inc., 1-6-1-438 Otemachi, Chiyoda-ku, Tokyo, Japan 100-0004}

\author{Takashi Onaka\altaffilmark{1,3}}
\altaffiltext{3}{Department of Physics, Meisei University, 2-1-1 Hodokubo, Hino, Tokyo, Japan 191-0042}

\author{Fr\'ed\'eric Galliano\altaffilmark{4,5}}
\altaffiltext{4}{IRFU, CEA, Universit\'e Paris-Saclay, F-91191 Gif-sur-Yvette, France}
\altaffiltext{5}{Universit\'e Paris-Diderot, AIM, Sorbonne Paris Cit\'e, CEA, CNRS, F-91191 Gif-sur-Yvette, France}

\author{Ronin Wu\altaffilmark{5,6}}
\altaffiltext{6}{Iris.ai, Edelgranveien 28 Bekkestua, 1356 Norway}

\author{Yasuo Doi\altaffilmark{7}}
\altaffiltext{7}{Department of Earth Science and Astronomy, University of Tokyo, Komaba 3-8-1, Meguro, Tokyo, Japan 153-0902}

\author{Hidehiro Kaneda\altaffilmark{8}}
\altaffiltext{8}{Nagoya University, Furo-cho, Chikusa-ku, Nagoya 464-8602, Japan}

\author{Daisuke Ishihara\altaffilmark{8}}

\author{Martin Giard\altaffilmark{9}}
\altaffiltext{9}{IRAP, Universit\'e de Toulouse, CNRS, CNES, 31028, Toulouse Cedex 4, France}

\email{abell@astron.s.u-tokyo.ac.jp}


\maketitle

\begin{abstract}
The anomalous microwave emission (AME) still lacks a conclusive explanation.
This excess of emission, roughly between 10 and 50~GHz, tends to defy attempts to explain it as synchrotron or free-free emission.
The overlap with frequencies important for cosmic microwave background explorations, combined with a strong correlation with interstellar dust, drive cross-disciplinary collaboration between interstellar medium and observational cosmology.
The apparent relationship with dust has prompted a ``spinning dust'' hypothesis.
The typical peak frequency range of the AME profile implicates spinning grains on the order of ~1~nm.
This points to polycyclic aromatic hydrocarbons (PAHs).
We use data from the AKARI/Infrared Camera (IRC), due to its thorough PAH-band coverage, to compare AME from the Planck Collaboration astrophysical component separation product with infrared dust emission in the $\lambda$~Orionis AME-prominent region.
We look also at infrared dust emission from other mid IR and far-IR bands.
The results and discussion contained here apply to an angular scale of approximately 1$^{\circ}$.
We find that certainly dust mass correlates with AME, and that PAH-related emission in the AKARI/IRC 9~$\mu$m band correlates slightly more strongly.
Using hierarchical Bayesian inference and full dust spectral energy distribution (SED) modeling we argue that AME in $\lambda$~Orionis correlates more strongly with PAH mass than with total dust mass, lending support for a spinning PAH hypothesis within this region.
We emphasize that future efforts to understand AME should focus on individual regions, and a detailed comparison of the PAH features with the variation of the AME SED.
\end{abstract}


\section{Introduction}
\label{sec:intro}

    Multi-wavelength observations have opened up a new discipline: the disentanglement of interstellar dust emission from non-dust emission components.
    A major challenge within this discipline is separating temperature fluctuations in the cosmic microwave background (CMB) from the microwave extent of thermal dust emission.

    The difficulty of decomposing the microwave sky into galactic interstellar medium (ISM), extragalactic, and CMB temperature fluctuation components has brought the detailed decomposition of the microwave-radio regime of ISM to the forefront Planck-based research \citep{planckEarly11I,planck2013I,planck2015I}.

    In our efforts to decompose and understand galactic microwave emission itself, there remains an antagonist.
    Galactic foregrounds had been broken down into 3 dominant components: free-free emission from ionized regions, synchrotron emission generated by relativistic electrons moving around the Milky Way's magnetic field, and the microwave extent of thermal dust emission \citep{wmap03b, leach08, planckXII}.
    Deviations from this understanding began to appear in the early 1990s, with efforts by \citet{kogut96, deoliveiracosta97} to carefully investigate the CMB.
    They had found a surprisingly good correlation between the microwave sky (free-free and synchrotron) and the far-infrared (FIR) dust emission at large angular scales.
    \citet{leitch97} was actually the first author to point to an excess of emission with respect to what could be expected from free-free and synchrotron emission, and to refer to this question as anomalous microwave emission (AME).
    Then \citet{deoliveiracosta98} published a detailed analysis of 19~GHz full sky survey data with previous Differential Microwave Radiometers, Far-InfraRed Absolute Spectrophotometer and radio surveys which showed a microwave excess with respect to free-free and thermal dust emission.
    AME generally takes the form of an excess continuum emission source, having a peak somewhere between 10 to 50~GHz.

    This excess defies predictions for known microwave emission mechanisms.
    AME still lacks a concrete physical explanation.
    Also, the term itself can be a bit confusing, as the word ``anomalous'' tends to imply a localized outlier.

	Since its first detection in early microwave observations, AME has been found to be a widespread feature of the microwave Milky Way (see the review by \citealp{dickinson13r}, and the state-of-play of AME research by \citealp{dickinson18}).
    \citet{finkbeiner02} reported the first detection of a ``rising spectrum source at 8~to~10~GHz'' in an observation targeting galactic ISM clouds.
    \citet{deoliveiracosta02} further argued that this emission is in fact ``ubiquitous''.
    This has been fully confirmed by detailed analysis with the Planck satellite data which shows excess AME-like emission at 30~GHz throughout the galactic plane (\citealp[see][Fig.~12]{planckxx11}).

  	More recent works, employing observations by the Wilkinson Microwave Anisotropy Probe (WMAP), Spitzer Space Telescope, the latest infrared to microwave all-sky maps by Planck, and various ground based radio observations have strongly confirmed a relationship between interstellar dust emission and AME \citep{ysard10a,tibbs11,hensley16}.
    Exactly which physical mechanisms produce the AME however is still an open question.
    We are equally puzzled as to the chemical composition and morphology of the carrier(s).
    We also lack an all-sky constraint on the emissivity of the AME spectrum at frequencies short of the WMAP cut-off, around 23~GHz.
    The typical peak frequency of AME, in those cases where it is constrained, does give us a clue.

	From the observed spatial correlation between AME and dust emerged two major hypotheses.
    The first and more widely supported is rotational emission from spinning small dust grains which have permanent electric and/or magnetic dipoles.
    The rotation of interstellar grains was originally studied in detail in order to compute grain alignment, which could explain the polarization of interstellar light \citep{davis51}.
    Then some authors mentioned rotational emission as an efficient relaxation process for small particles \citep{draine79, rouan92}, some of which may be able to reach suprathermal rotational velocities \citep{spitzer79, purcell79}.

    \citet{draine99} proposed a second, but not mutually exclusive mechanism, commonly referred to as magnetic dust emission.
    In the magnetic dust case, thermal fluctuations in magnetization of grains with magnetic inclusions may contribute to AME.

    Though not widely accepted, another minor explanation for AME is discussed in \citet{jones09}. 
    They have suggested that the emissivity of dust, in the spectral range related to AME, could contain features caused by low temperature solid-state structural transitions. 
    To date however, the spinning dust hypothesis has been the more widely supported case, and is thus the focus of this work.

    The mechanism of spinning dust emission was first proposed in \citet{erickson57} and \citet{hoyle70}, with further discussion in \citet{ferrara94}. 
    \citet{draine98b} give the earliest thorough theoretical prediction of a spinning dust spectrum. 
    A decade later \citet{ali-haimoud09} contributed substantial updates, expanded modeling of grain excitation mechanisms and adoption of an updated grain size distribution by \citet{weingartner01}.
    Modeled spectra for potential candidate carriers of spinning dust have since appeared in the literature \citep{ysard10a, draine13, ali-haimoud14, hoang16a}. 
    The photometric signature of AME is frequently interpreted via spinning dust parameters \citep{ysard11,ali-haimoud10}.

    \citet{draine98b}, gives the expected rotational frequency of spinning dust oscillators $\omega$,  as follows:
	        \begin{equation}
	        \frac{\omega_T}{2\pi} =
	        \langle\nu^2\rangle^{1/2}
	        \approx 5.60\times 10^9 a^{-5/2}\xi^{-1/2}T^{1/2}~~~{\rm Hz},
	        \label{eq:nurms}
	        \end{equation}
	   where $T$ is the gas temperature in K, $a$ is the grain size in nm, and $\xi$ represents the deviation from a spherical moment of inertia. 
	Assuming a typical peak frequency of the AME of \textasciitilde{}20~GHz, a gas temperature of 100~K, and roughly spherical grains, Eq.~\ref{eq:nurms} implies an oscillator size of approximately 1~nm.
     Based on the knowledge available at the time and the implied size range, \citet{draine98b} proposed polycyclic aromatic hydrocarbons (PAHs) as potential carriers of AME. 
     \citet{ysard10a} introduced the first model of a spinning dust spectrum based on rotational emission from PAHs.

    The major observational indication for PAHs in the ISM is the collection of Unidentified Infrared Band(s) from \textasciitilde{}3 to 17~$\mu$m. 
    These were first reported at 8 to 13~$\mu$m in planetary nebulae and HII regions by \citet{gillet73, gillett75} in ground-based observations. 
    \citet{leger84} offered the earliest suggestion that these features may come from PAHs. 
    Balloon-based observation by \citet{giard88, giard89, giard94} confirmed that the 3.3~$\mu$m feature pervades throughout the Galactic plane. 
    Space-based spectroscopy with the Infrared Telescope in Space (IRTS) \citep{murakami96} and Infrared Space Observatory \citep{kessler96} enabled confirmation by \citet{onaka96} and \citet{mattila96} that the mid-infrared (MIR) UIB features are not limited to a few objects, but are present even in the diffuse galactic ISM. 
    \citet{dwek97} even reported that photometric excess in the Cosmic Background Explorer/Diffuse Infrared Background Experiment (DIRBE) 12~$\mu$m band may be caused by emission from PAH, a possibility which had been considered earlier by \citet{allamandola85,puget85}. 
    The review by \citet{tielens08}, considering observational, theoretical, and experimental works on the subject, strongly supports not only the presence, but the ubiquity of large PAHs in the ISM. 
    \citet{andrews15} argue for the existence of a dominant ``grandPAH'' class, containing 20 to 30 PAH species.  
    The terms ``PAH features'' and ``UIBs'' are often used interchangeably in the literature, since there is yet no firm identification of specific molecular PAH species that are able to reproduce the observed infrared bands. 
    PAH-like amalgams containing aromatic structures have been incorporated into dust mixture spectral energy distribtion (SED) models over the last two decades \citep{drli01, drli07, hony01, dustem11, galliano11, jones13, jones17}. 
    Examples include mixed aromatic-aliphatic organic nanoparticles (MAONs) per \citet{kwok11} and quenched carbonaceous composites (QCCs) described by \citet{sakata84}. 
    A more in-depth, up-to-date discussion of PAH studies is contained in the review by \citet{galliano18b}.

    In the following sections, we will define PAHs as a general class of aromatic-ring-containing molecules, and not explicitly to specific PAH species such as benzene, corronene, or corranulene, etc. From this point we will also abandon the term ``UIBs'', and use ``PAH features'' instead.

   	In any case, the PAH class of molecules are the only spinning dust candidate so far which show both: \\
   	1) Evidence of abundance in the ISM at IR wavelengths, and \\
   	2) A predicted range of dipole moments (on order of 1~debye), to produce the observed AME signature \citep{draine98b, lovas05, thorwirth07}.

   	However, it should be noted that although nanosilicates have not yet been detected in the ISM, \citet{hensley17a} propose that an upper bound on the abundance of nanosilicates by \citet{li01}, based on IRTS observations by \citet{onaka96}, allow such small spinning grains to be composed primarily of silicates. 
   	This followed an earlier claim by \citet{hensley16} that the absence of a conclusive PAH-AME link suggested the possibility of AME-from-nanosilicates. 
   	The MIR nanosilicate emission, in such a scenario, is thus implied to be hidden by the complex PAH spectrum.

   	While neither nanosilicates nor any particular species of PAH have been conclusively identified in the ISM, there is far more empirical evidence for PAH-like dust than for nanosilicates.
   	To date, there have been no reports of suspected spinning dust emission from any region thought to be PAH-poor. 
   	The observational evidence that AME correlates with thermal dust emission, goes hand-in-hand with the presence of PAH emission. 
   	Several works have set out to answer the question of whether thermal dust emission or PAH emission correlate better with the AME.

	  Archival all-sky AME data products exclusively assume spinning dust SED templates. 
	  Both WMAP and Planck used a base template with 30~GHz peak frequency, and an assumed cold neutral medium environment. 
	  The {\tt SpDust} \citep{ali-haimoud10, silsbee11} spinning dust SED model code\footnote{Available at: http://pages.jh.edu/~yalihai1/spdust/spdust.html} has become a common tool when interpreting microwave foreground emission.

 	In the spinning dust model, there are several possible excitation factors for spinning dust. 
 	For the grains to have rotational velocities high enough to create the observed AME, they must be subject to strong excitation mechanisms. 
 	The dominant factors that would be giving grains their spin, are broken down by \citet{draine11} into basically two categories: 1) Collisional excitation. 
 	2) Radiative excitation, the sum of which could lead to sufficient rotational velocities for sufficiently small grains. 
 	However the extent of excitation will depend on environmental conditions. More frequent encounters with ions and atoms will occur in denser regions, so long as the density is not high enough to coagulate the small grains onto big-grains (BGs). 
 	
 	Likewise, more excitation due to photon emission will occur with increasing insterstellar radiation field (ISRF) strength \citep{ali-haimoud09, ali-haimoud14}.

    One of the strongest potential excitation mechanisms listed in \citet{draine11} is that of negatively charged grains interacting with ions. 
    Thus not only must we consider environmental factors, grain composition and size, but also the ionization state of the carriers. 
    The dependence of the observed AME on ISM density is modeled by \citet{ali-haimoud10}.

    We explore the case that the AME signature arises from spinning dust emission. 
    If the AME is carried by spinning dust, the carrier should be small enough that it can be rotationally excited to frequencies in the range of 10-50~GHz, and must have a permanent electric dipole. Within contemporary dust SED models, only the PAH family of molecules, or nanoscale amorphous carbon dust fit these criteria. 
    PAHs with a permanent electric dipole can emit rotationally.

    The overall pattern among large-scale studies is that the dust-tracing photometric bands correlate with the AME (and each other) to first-order \citep{dickinson13r}. 
    On an all-sky, pixel-by-pixel basis, at 1$^{\circ}$ angular resolution, \citet{ysard10b} find that 12~$\mu$m emission, via Infrared Astronomical Satellite (IRAS), correlates slightly more strongly with AME (via WMAP) than with 100~$\mu$m emission. 
    They also find that normalizing the IR intensity by the ISRF improves both correlations. 
    They interpret this finding as evidence that AME is related to dust, and more closely related to stochastically heated very small grains (VSGs) and PAHs--- which are traced by 12~$\mu$m emission. 
    The improvement of the correlations after normalizing by the ISRF is expected, as long as the 12~$\mu$m band is dominated by stochastic emission from PAHs, in other words:
            \begin{equation}
              I_{12~\mu{}m} \propto{} UN_{PAH},
            \end{equation}
    where $N_{PAH}$ is the column density of emitting PAHs \citep{onaka00}, and $U$ is the strength of the ISRF normalized to that of the solar neighborhood as provided in \citet{mathis83}, where $U = 1$ corresponds to approximately $1.41\times{}10^{-3} {\mathrm{erg s^{-1} cm^{-2}}}$. 
    Throughout this work we will use $U$ to indicate ISRF intensity.

    Such a relationship is expected, if we make two major assumptions. 
    Regarding PAHs, they must be small enough and their heat capacities low enough, such that their heating is indeed stochastic-- a single UV photon is absorbed, immediately followed by the emission of many IR photons. 
    The second assumption, is that the spectral shape of the observed MIR and FIR dust emission will vary according to the excitation conditions, while the PAH spectrum will not \citep[see][Fig.~13]{li01}. 
    Under such conditions, the radiation field would change only the intensity of PAH features. 
    Thus \citet{ysard10b} implies that $I_{12~\mu{}m}/U$ is giving us a measure of the column density of spinning dust.

    In a similar work however \citet{hensley16} report a lack of support for the spinning PAH hypothesis. 
    Finding that fluctuations in the ratio of PAH-dominated 12~$\mu$m emission, via the Wide-field Infrared Survey Explorer to dust radiance, $R$ via Planck, do not correlate with the ratio of AME intensity to $R$, they conclude that the AME is not likely to come from PAHs. 
    In terms of emission intensity however, their findings are consistent with \citet{ysard10b} in that $I_{12~\mu{}m}$ correlates well with $I_{\rm AME}$. 
    Thus there remains an open question as to what the actual carrier of the AME is.

    The story is no more clear when looking at the average properties of individual regions.
    \citet{planckXV} find that among 22 high-confidence ``AME regions'' (galactic clouds such as the $\rho$~Ophiuchus cloud and the Perseus molecular cloud complex) AME vs. 12~$\mu$m  shows a marginally weaker correlation than AME vs. 100~$\mu$m (via IRAS). 
    \citet{tibbs11} examined the AME-prominent Perseus Molecular Cloud complex, finding that while there is no clear evidence of a PAH-AME correlation, they do find a slight correlation between AME and $U$.

    In general there appear to be two types of studies: those that examine AME across the sky and attempt to answer, in a global sense, ``What causes AME?''. 
    Then we have targeted studies which attempt to describe the AME within particular regions. 
    Statistical works addressing AME-at-large have been largely inconclusive in determining whether or not spinning PAHs contribute to the AME. 
    If there are indeed local environmental dependencies on the AME, these may be blurred by an inherent mixing of environments along a given line of sight in an all-sky survey.
    We may expect that less localized studies could miss subtle trends between the AME and the IR dust SED. 
    This may be one reason why all-sky AME works so far have not been more conclusive.

    More recently, \citet{planck15XXV} noted that although there are significant degeneracies between the free-free, synchrotron, and spinning dust component maps across the sky, there is a particular region of the sky which shows a high AME emissivity vs. the Planck 545~GHz cold-dust-tracing emission. 
    The authors suggested $\lambda$~Orionis as a strong candidate for further AME investigation, due not only to the high AME emissivity, but also the region's interesting morphology.
    
    The $\lambda$~Orionis molecular ring is an approximately 10$^{\circ}$ wide ring-shaped structure in IR emission, surrounding the $\lambda$~Orionis O-type star,  
    \citet{maddalena86,maddalena87, zhang89} noted a ring of dust emission, and molecular cloud being pushed out by the $\lambda$~Orionis Association of B-type stars. 
    The inner ring traces the boundary of a well-known PDR, which surrounds an HII region ionized by $\lambda$Ori itself and its OB associates \citep{murdin77, ochsendorf15}. 
    The ring shape itself is thought to originate from a supernova, or perhaps combined effects of the entire star formation history of the $\lambda$~Orionis Association, including the formation of its surrounding HII region \citep{cunha96, aran09}and PDR. 
    The region is known to host several young stellar and protostellar objects \citep{koenig15}.
    
    This transition of environments makes $\lambda$~Orionis an interesting target for AME study, as we can make a more robust test of the correlation strength between AME and PAH emission. 
    The free-free emission itself shows a very different spatial morphology from the AME and dust emission, such that the AME here appears to be reliably separated from other microwave components.
    Confusion with free-free emission is one issue which complicated analysis of other HII regions such as RCW 175 \citep{tibbs12b}, where the free-free prominent.
    Due to the large angular size, we can visibly separate the characteristic environments of $\lambda$~Orionis even in the 1$^{\circ}$ full-width at half-max (FWHM) resolution Planck Collaboration (PC) AME map.
    The angular distance from the galactic plane also reduces confusion with background sources.  
    
    Thus in the present work we explore $\lambda$~Orionis further,looking at its IR dust SED across more wavebands than have been examined in \citet{planck15XXV} and compare this to the AME map introduced in \citet{planck15X} (hereafter, PC15X).
    
    In the following sections we will explore a potential connection between emission from PAHs and AME within $\lambda$~Orionis. 
    The focus will be on the AKARI/IRC 9~$\mu$m band which allows a more complete study of the PAH features than has been possible with IRAS or WISE data. 
    In Sect.~\ref{sec:datasources} we will detail the data sources used, from the infrared maps tracing dust and PAH emission to the microwave component separation maps produced by the PC which give an estimate of the AME. 
    Sect.~\ref{sec:dataproc} details how the data for $\lambda$~Orionis are extracted from the respective data sources, processed, and otherwise prepared for a side-by-side analysis. 
    In order to make a physical interpretation of the data, Sect.~\ref{sec:analysis} details both a correlation test of which infrared waveband correlates best with AME as well as a full dust SED fitting based on a hierarchical Bayesian (HB), the results of which are shown in \Cref{sec:results}. 
    To assess which component of dust shows a stronger relationship with the Planck AME data, we show also the probability density function (PDF) of correlation tests between AME and three dust parameters: mass of PAHs, total mass of dust, and mass of ionized PAHs.
    In Sect.~\ref{sec:discussion} we provide the first conclusive argument for the AME-carried-by-PAHs, based on SED fitting, for a particular region of the sky. 
    This work represents the first time that either AKARI/IRC 9~$\mu$m data or the hierarchical Bayesian analysis (HB) approach by \citet{galliano18a} have been used for AME investigation.
    Full dust SED fitting as well has not been applied to the $\lambda$~Orionis region.

  \section{Data sources}
  \label{sec:datasources}
\begin{figure*}
       \begin{center}
         \includegraphics[width=\textwidth,trim={2cm 0.5cm 3.0cm 1cm},clip]{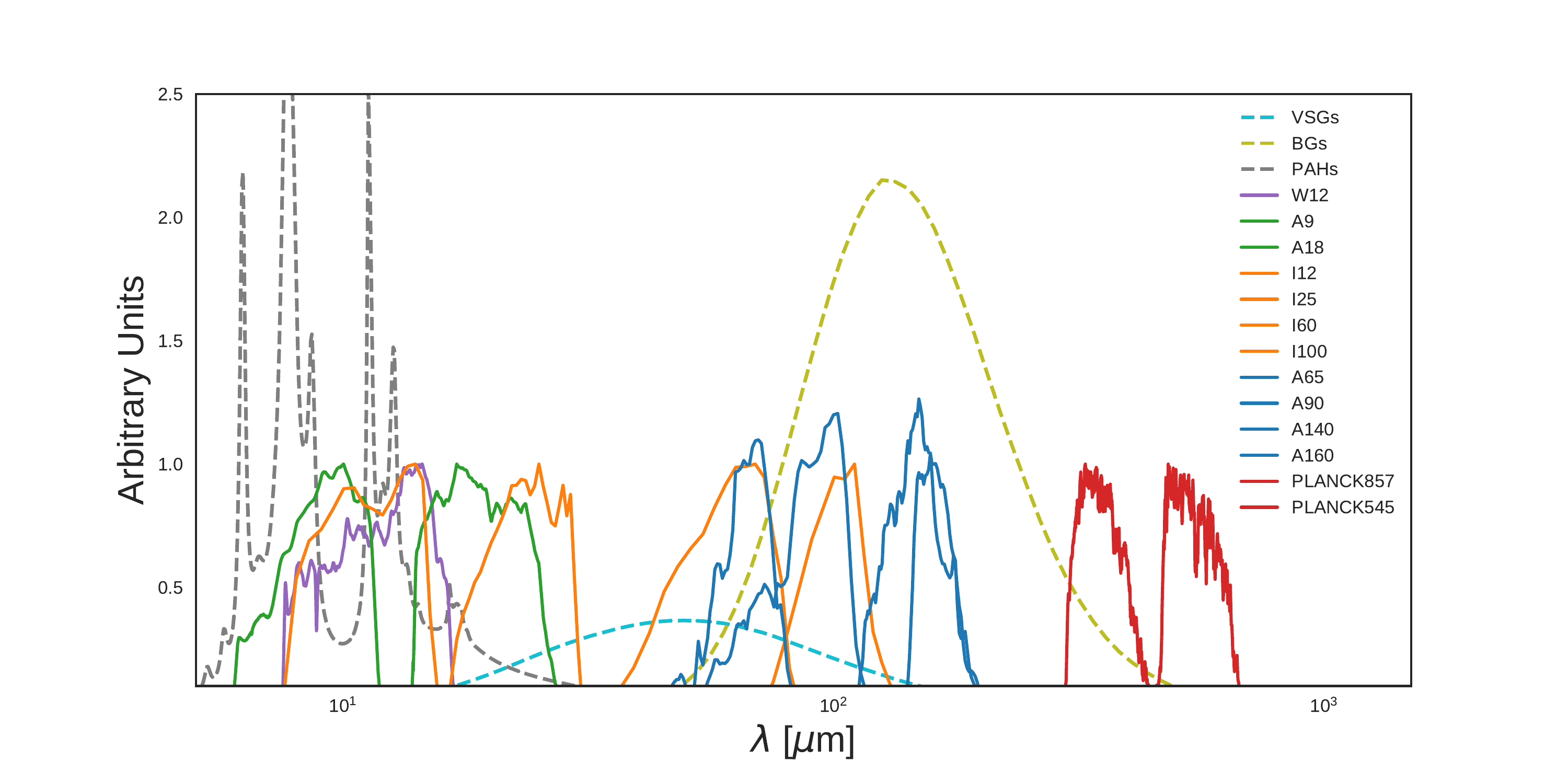}
       \end{center}
       \caption{Relative spectral response curves of the bands used in this study: AKARI/IRC 9 and 18 $\mu$m bands (green); AKARI/FIS 65, 90, 140 and 160~$\mu$m (orange); IRAS 12, 25, 60 and 100~$\mu$m (orange); and the Planck/HFI 857~GHz and 545~GHz bands (red). 
       The WISE 12~$\mu$m band is shown only for comparison (solid lavender), and is not used in this study. Expected dust emission components, assuming the dust SED model by \citet{dustem11}, at an ISRF strength of $U=1$, are also shown. 
       The components are summarized as emission from BGs (dashed yellow line), emission from VSGs (dashed blue line), and emission from PAHs (dashed grey line).}
       \label{fig:Filter_coverage_example_full}
     \end{figure*}

    In order to cover the large solid angle subtended by $\lambda$~Orionis region, across multiple MIR to FIR wavelengths, we extract data from all-sky surveys. 
    All of the maps utilized are photometric-band infrared maps, except for the AME data, which is an all-sky component separation analysis product, from the PC efforts to separate galactic foregrounds from the CMB PC15X. 
    Table~\ref{tab:data} summarizes the observational data used in this work.
               \begin{table*}
                 \caption{Table of data sources used in this work.}
                   \begin{tabular}{lrrrrr}
                   \hline\hline
                   Instrument & Central    & FWHM & Calibration & Reference\footnotemark[$*$] & Coverage \\
                              & Wavelength &            & Uncertainty &           & Contribution \\
                   \hline
                   AKARI/IRC & 9~$\mu$m  &  \textasciitilde{}10$"$ & \textless 10\%   & [1] & VSGs; 6.2, 7.7, 8.6 \\
                   & & & & & and 11.2~$\mu$m PAH \\
                   AKARI/IRC & 18~$\mu$m & \textasciitilde{}10$"$  & \textless 10\%     & [1] & VSGs \\
                   AKARI/FIS & 65~$\mu$m  & 63$"$ &   \textless 10\% & [2,3] & VSGs, TEGs\footnotemark[$a$] \\
                   AKARI/FIS & 90~$\mu$m  & 78$"$ &   \textless 10\%   & [2,3] & TEGs\\
                   AKARI/FIS & 140~$\mu$m & 88$"$ &   \textless 10\%   & [2,3] & TEGs\\
                   AKARI/FIS & 160~$\mu$m & 88$"$ &   \textless 10\%   & [2,3] & TEGs\\
                   COBE/DIRBE & 160~$\mu$m & 88$"$ &  \textless 10\%   & [2,3] & TEGs\\
                   IRAS/IRIS & 12~$\mu$m   & 4.0$'$ & \textless 5.1\%       & [4] & VSGs, 8.6, 11.2, 12.7~$\mu$m PAH features \\
                   IRAS/IRIS & 25~$\mu$m   & 4.0$'$ & \textless 15.1\%      & [4] & VSGs\\
                   IRAS/IRIS & 60~$\mu$m   & 4.2$'$ & \textless 10.4\%      & [4] & VSGs, TEGs\\
                   IRAS/IRIS & 100~$\mu$m  & 4.5$'$ & \textless 13.5\%       & [4] & TEGs \\
                   Planck/HFI           & 345~$\mu$m     & 4.7$'$  & & [5] & TEGs \\
                   Planck/HFI           & 550~$\mu$m     & 4.3$'$  & & [5] & TEGs \\
                   $H\alpha{}$          & 658.5~nm       & 36$'$   & & [6] & free-free tracer\\
                   Planck $AME_{\rm var}$ (PR2) & 22.8~GHz   & 60$'$   & & [7] & variable peak frequency AME \\
                   Planck $AME_{\rm fix}$ (PR2) & 41.0~GHz   & 60$'$   & & [7]                             & fixed peak frequency AME \\
                   \hline
                 \end{tabular}
                 \begin{tablenotes}
                   \item{$a$} Thermal Equilibrium Grains
                   \item{$*$} [1]~\citet{ishihara10}; [2]~\citet{doi15}; [3]~\citet{takita16}; [4]~\citet{iris05}; [5]~\citet{hfi14viii}; [6]~\citet{finkbeiner03}; [7]~PC15X
                 \end{tablenotes}

                 \label{tab:data}
               \end{table*}
    In total, we employ data from 12 photometric bands, spanning the wavelength range of 6.9~$\mu$m to 550~$\mu$m as shown in Fig.~\ref{fig:Filter_coverage_example_full}. 
    The following sections give the details of the observational data from each instrument as well as of the parameter maps provided in PC15X.\footnote{Planck bands are named according to their central frequency, not wavelength.}
    From this point we will mostly use abbreviations to refer to the different bands, as follows: 'A' indicates AKARI; 'D', DIRBE; 'I', IRAS, and 'P' Planck. 
    The number after each letter indicates the band central nominal wavelength in microns (or frequency in GHz, in the case of the Planck bands).

 \subsection{AKARI}
 \label{sec:AKARI}

   The AKARI IR space telescope \citep{akari07} enhanced our view of the infrared sky from the mid to far infrared via two instruments, the Infrared Camera (IRC) \citep{irc07} and the Far-Infrared Surveyor \citep{kawada07}. 
   The
   MIR wide-band filter positions at 9~and~18~$\mu$m compliment those of COBE, IRAS and WISE, providing the community with an overall denser photometric sampling of all-sky MIR emission. 
   The FIR bands provide the highest resolution all-sky surveys yet for 60~to~180~$\mu$m range. 
  \subsubsection{
  Infrared Camera
  }
   IRC provided us with both spectroscopic and photometric data from the near to mid-infrared. 
   In this work, we utilize data from the bands centered at 9 and 18~$\mu$m, created during the IRC's fast-scanning phase. 
   We utilize the most recent version of the IRC data (Ishihara, et al., in prep.). This version has had an updated model of the zodiacal light, fitted and subtracted. 
   The details of the improved zodi-model, which offers an improvement over that used for the IRAS all-sky maps, are given in \citet{kondo16}.
  \subsubsection{
  Far-infrared surveyor)
  }
    FIS gives us photometric data around the peak of the typical thermal dust SED. 
    FIS was equipped with four wavebands: two narrow bands centered at 65~$\mu$m and at 160~$\mu$m, and two wide bands at 90~$\mu$m and at 140~$\mu$m. 
    An all-sky survey was carried out at each band \citep{kawada07}, and the processed maps, in which the major zodiacal emission has been subtracted, have been publicly released \citep{doi15}.

  \subsection{Planck Observatory High Frequency Instrument }
    The High Frequency Instrument (HFI) all-sky maps, spanning 100 to 857~GHz \citep{hfi14viii} help constrain the 
    FIR dust emissivity. 
    This study utilizes the 857~GHz (345~$\mu$m) and 545~GHz (550~$\mu$m) bands.

  \subsection{Infrared Astronomical Satellite (IRAS)}
    Data from the IRAS \citep{iras84} all-sky surveys are used to supplement the similarly-centered AKARI photometric bands. 
    The I12 band is similar to the A9 band in terms of the sky coverage in that both surveys are heavily dominated by zodiacal light. 
    Also, while both cover emission from PAH features, I12 is more sensitive to warm dust emission. 
    We use the Improved Reprocessing of the IRAS Surveys (IRIS) \citep{iris05}, which have undergone a zodiacal-light removal. 
    We do not utilize the WISE data because such high resolution was not required; we are conducting our analysis at 1$^{\circ}$ resolution in order to match that of the AME data modeled by PC15X.
    
    \subsection{Planck COMMANDER AME Parameter Maps}
    \label{sec:PCmaps}
       We utilize the AME map from the PC COMMANDER-Ruler astrophysical component separation described in PC15X, a part of the PC
        Public Data Release 2 (PR2) \citep{planck2015I}. 
        These contain estimates of known microwave foreground components (free-free, synchrotron, thermal dust emission contributions to the Planck photometric bands). 
        The effective beam size of these maps is given by PR2 as 1$^{\circ}$.

        The ``AME component map'' presumes that AME originates from spinning dust. 
        While acknowledging that such a decomposition lacks a strong physical interpretation, PC15X break the AME into two components: a spatially varying peak frequency component, $AME_{\rm var}$, and a spatially constant peak frequency component, $AME_{\rm fix}$.

        We stress that while this is the most careful all-sky attempt to isolate the AME to date, the constraints of the frequency peak are still heavily lacking. 
        In general, $AME_{\rm var}$ is the dominant component. 
        The intensities given by PC are evaluated at reference frequencies: 22.8~GHz for $AME_{\rm var}$, 41~GHz for $AME_{\rm fix}$ ). 
        They are not fitted peak intensities of the spinning dust SED model.

         Because of the phenomenological approach of the AME fitting method, the PC15X authors themselves suggest caution in deriving conclusions from comparisons with the COMMANDER AME map. 
         However it is the most thorough all-sky component separation currently available, and has not been well analyzed relative to the full wavelength range of available IR all-sky maps.
\section{
PAH feature coverage
}
\begin{figure*}
                         \begin{center}
                         \includegraphics[width=\textwidth]{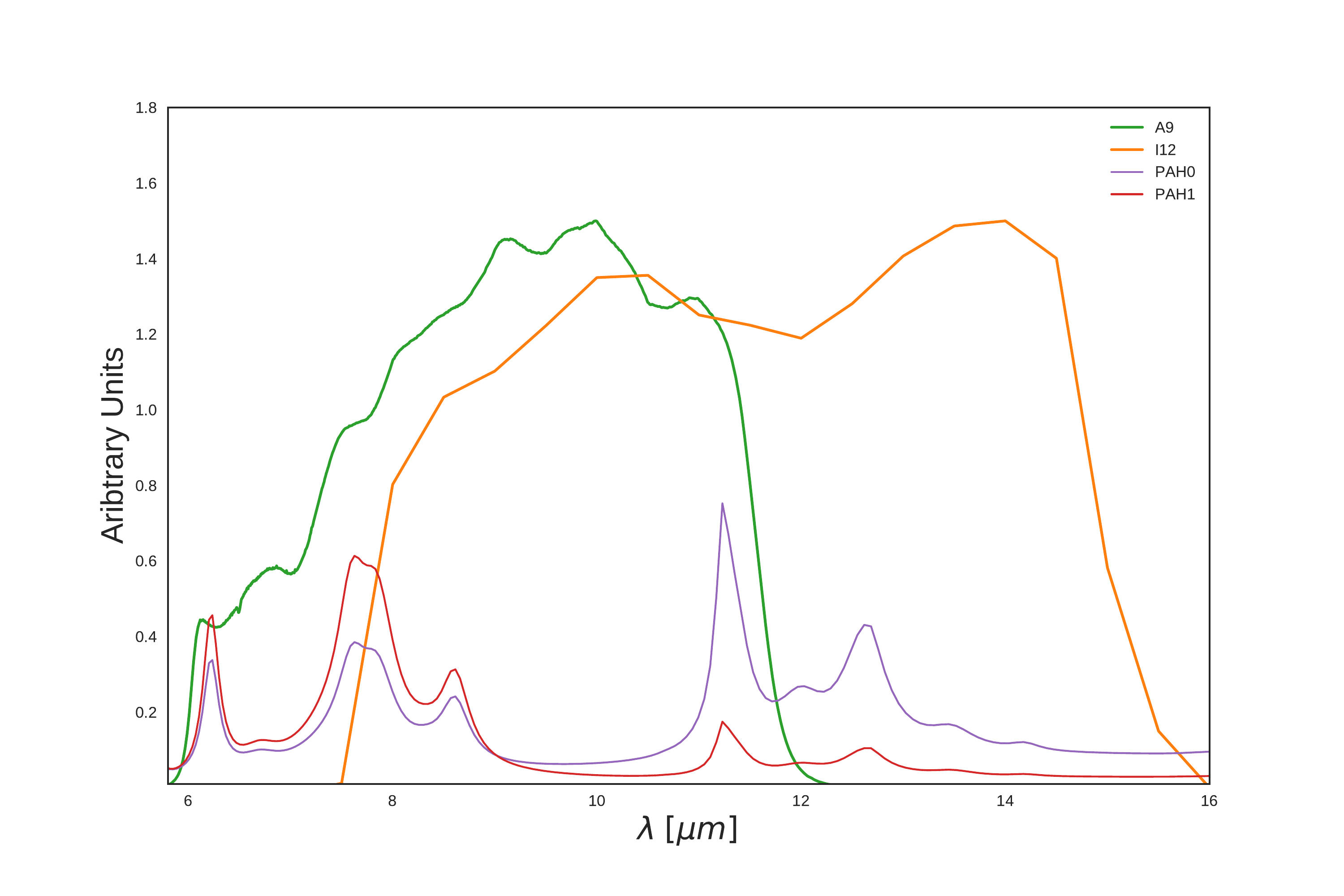}
                         \end{center}
                         \caption{The I12 (orange) and A9 (green) filters coverage of modeled ionized (PAH1, red) and neutral (PAH0, purple) components of PAH features by \protect{\citet{dustem11}}. 
                         The difference in the PAH feature coverage mainly comes from the 6.2~$\mu$m and the 7.7~$\mu$m features.}
                         \label{fig:Filter_coverage_example_PAH}
                       \end{figure*}

\begin{figure*}
                          \begin{center}
                          \includegraphics[width=\textwidth]{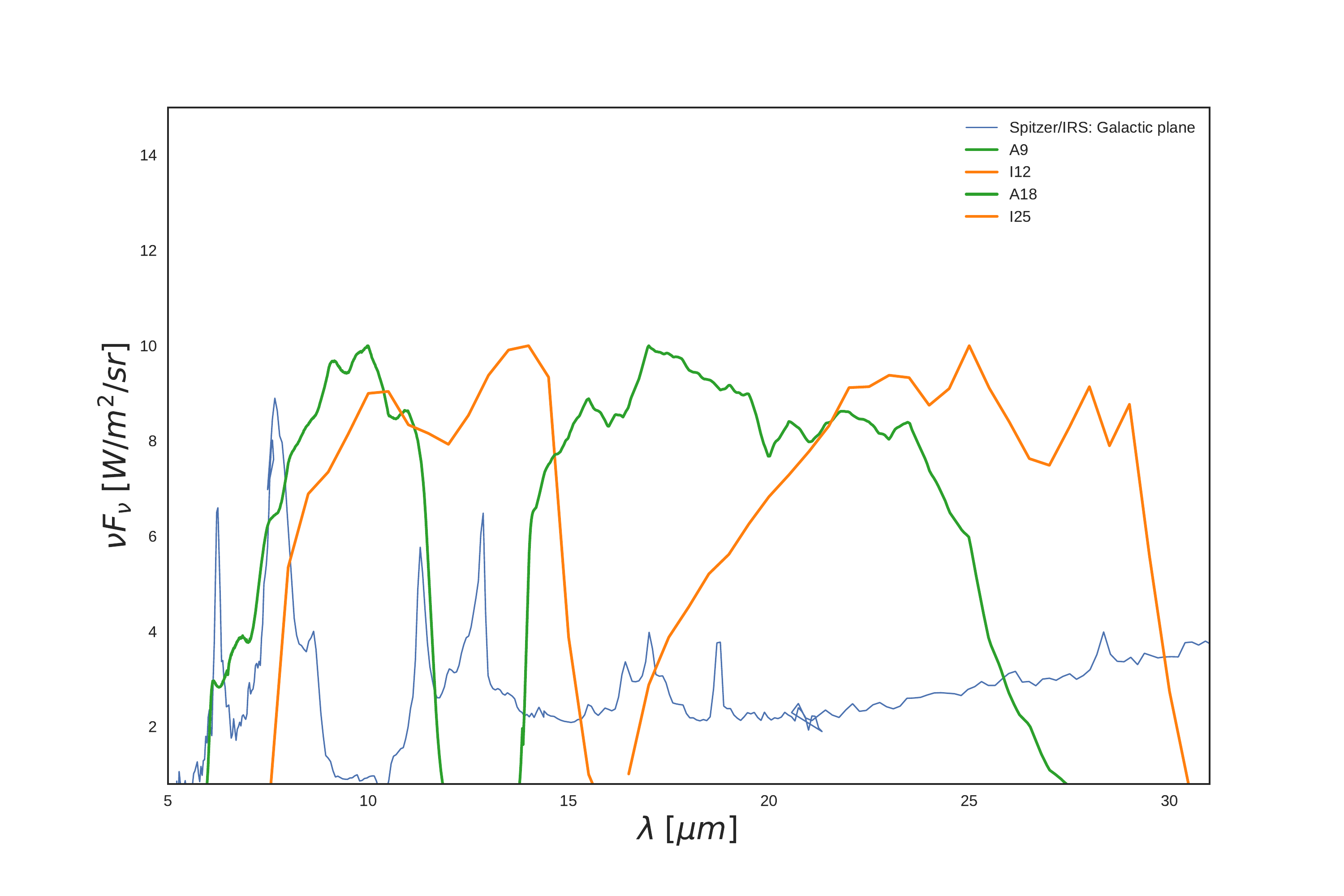}
                          \end{center}
                          \caption{Coverage of MIR wavelengths by the filters used in this work. 
                          A Spitzer/IRS spectrum (see AOT4119040) of the galactic plane (thin blue line) demonstrates how IRC and IRAS photometric bands trace these features on an all-sky basis \protect{\citep{ishihara07}}. 
                          Strong PAH features overlap with the A9 and I12, while the A18 and I25 bands only trace much weaker features. }
                          \label{fig:Filter_coverage_example_MIR}
                        \end{figure*}

    The A9 all-sky map demonstrates the abundance of the PAH bands carrier in the Milky Way \citep{ishihara10}. 
    Fig.~\ref{fig:Filter_coverage_example_PAH} shows the coverage of the PAH features (from both ionized and neutral PAH components), as they are theoretically determined in the model of ISM IR cirrus emission, by \citet{dustem11}.
    The A9 band covers ionized PAH features at 6.2, 7.7, and the C-H bond feature at 8.6~$\mu$m; as well as the 11.2~$\mu$m feature from neutral PAHs, across the entire sky \citep{irc07}. 
    In particular, A9 is the only all-sky survey sensitive to the 6.2 and 7.7~$\mu$m features. 
    The I12 band covers the 11.2 and 8.6~$\mu$m features, and the similarly-shaped W12 band covers primarily the 11.2~$\mu{}$m feature but do not cover the 7.7~$\mu{}$m completely. 
    According to the distribution of PAH features across the response filters in Fig.~\ref{fig:Filter_coverage_example_PAH}, and referring to the various dust components in Fig.~\ref{fig:Filter_coverage_example_full} it is also expected that the A9 band is the most dominated by PAH emission even with increasing $U$. 
    This may seem counter-intuitive, since, as described in Sect.~\ref{sec:intro}, the PAH spectral shape does not show a temperature variation.
    However as $U$ increases, the shorter (\textasciitilde{}10~$\mu$m) extent of thermal dust emission and emission from VSGs encroach on I12 sooner than A9, diluting emission from PAHs.
    In some ionized regions, I12 may also include non-significant contributions from the [NeII] line at 12.8~$\mu$m. 
    Fig.~\ref{fig:Filter_coverage_example_MIR} demonstrates an example observational galactic cirrus spectrum in the MIR, the Spitzer/Infrared Spectrograph \citep{houck04} data, along with filters for all of the MIR bands used in this study.
    The other MIR bands, A18 and I25, do not cover strong PAH features and are expected to be dominated rather by emission from VSGs, as was indicated in Fig.~\ref{fig:Filter_coverage_example_full}.

    \subsection{
    PAH ionization
    }
     Fig.~\ref{fig:Filter_coverage_example_PAH} indicates that expected emission from ionized PAHs may preferentially contribute to the A9 band, even though both I12 and A9 cover ionized and neutral features.
A SED model calculation using the dust emission and extinction (DUSTEM) \citep{dustem11} supports that for Galactic cirrus ISM conditions, emission detected by the A9 has a higher contribution from charged PAHs than the I12 band.

Both bands include contributions from charged and neutral PAHs emission.  However, the relative contribution from charged PAHs emission is higher for A9.  Thus the ratio of the intensities in these two bands could trace the fraction of charged PAHs for a given line of sight.
     
\begin{figure*}
                         \begin{center} \includegraphics[width=\textwidth]{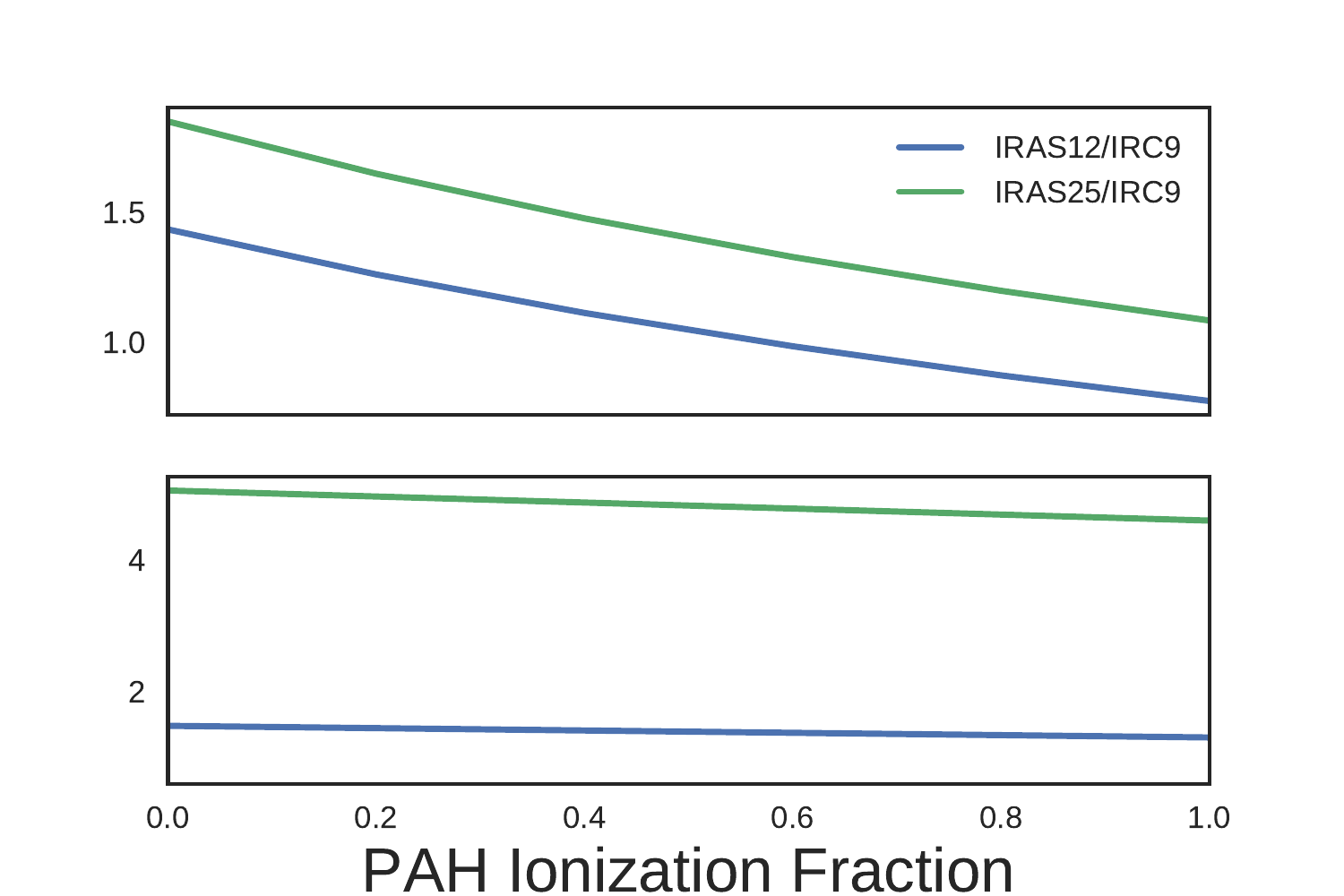}
                         \end{center}
                         \caption{Ionization fraction of PAHs vs. band ratios of I12, I25, to A9, for two values of $U$: Top: $U = 1$, Bottom: $U = 10$. These ratios are determined by assuming the SED template of \protect{\citet{dustem11}} }
                         \label{fig:band-ratio-multiple}
                     \end{figure*}

     Estimating the extent of this effect, Fig.~\ref{fig:band-ratio-multiple} gives the results of a calculation of I12/A9 and I25/A9 band ratios for two values of $U$.

     These contributions remain relatively constant out to a $U$ of about 100, with the contribution from warm dust becoming a larger factor for the I12 band. 
     Thus, according to the DL01 template, A9 should have the highest contribution from PAHs out to extreme radiation fields. 
     This calculation is again based on \citet{dustem11}. 
     To the extent with which PAHs can endure harsh UV radiation, as PAHs are expected to evaporate in strong enough radiation fields \citep{allain96a,allain96b,bocchio12,pilleri12, pavlyuchenkov13}.
      It suggests that at least for $U\lesssim{}10$, the fraction of charged PAHs may be estimated as a function of the I12/A9 ratio.

\section{Data processing}
\label{sec:dataproc}
  As indicated in Sect.~\ref{sec:datasources}, we use 12 photometric all-sky maps. 
  For the IRC data (A9 and A18), we produce mosaics of $\lambda$~Orionis from the latest version of the individual tiles provided by the internal all-sky archive. 
  A9 and A18 images are produced by regridding the images with the {\tt Montage} software by National Aeronautics and Space Administration/Infrared Processing and Analysis Center. 
  \Cref{fig:lori_A9_mosaic_smooth,fig:lori_A18_mosaic_smooth} show high resolution mosaics of the A9 and A18 data before processing.

  Other data were obtained from public archived Hierarchical Equal Area Iso Latitude Pixelation of the Sphere (HEALPix) maps. 
  This includes data from IRAS, Planck, DIRBE, and AKARI/FIS. 
  For these data, we employ the {\tt healpix2wcs} functionality provided in the {\tt gnomdrizz} python package.\footnote{Available at \texttt{http://cade.irap.omp.eu/dokuwiki/doku.php?id=software}} 
  When the extraction and regridding are finished, all of the images --- those extracted from HEALPix maps, as well as the IRC data --- share a common FITS header having a pixel grid spacing equal to the average pixel width in the NSIDE 256 HEALPix scheme, or about \textasciitilde{0.25$^{\circ{}}$}. 
  After point-spread function (PSF) smoothing, described in detail below, the data also have a common \textasciitilde{1$^{\circ{}}$} FWHM effective PSF. 
  Although this means that the grid is oversampled, we choose this approach to preserve as much of the structure of the region as possible, even after masking bad pixels. 
  Another option would be to interpolate the bad pixels, and degrade the pixel scale to the PSF size. 
  However we choose the former method due to the relatively large number of bad pixels, especially around the regions affected by missing stripes. 

  These stripe patterns arise from two factors: they represent regions where the de-striping processing applied by \cite{takita16} failed, due to strong small-scale brightness variations in certain region.
  Also, to accommodate pointed observation requests during AKARI's all-sky scan, some regions were omitted from the all-sky scanning process.
  
  This means that in the analyses that follow, what is important is the relative trends between correlation coefficients, rather than their absolute values.

  \subsection{Point-source and artifact masking}
\begin{figure*}
        \begin{center}
          \includegraphics[width=\textwidth]{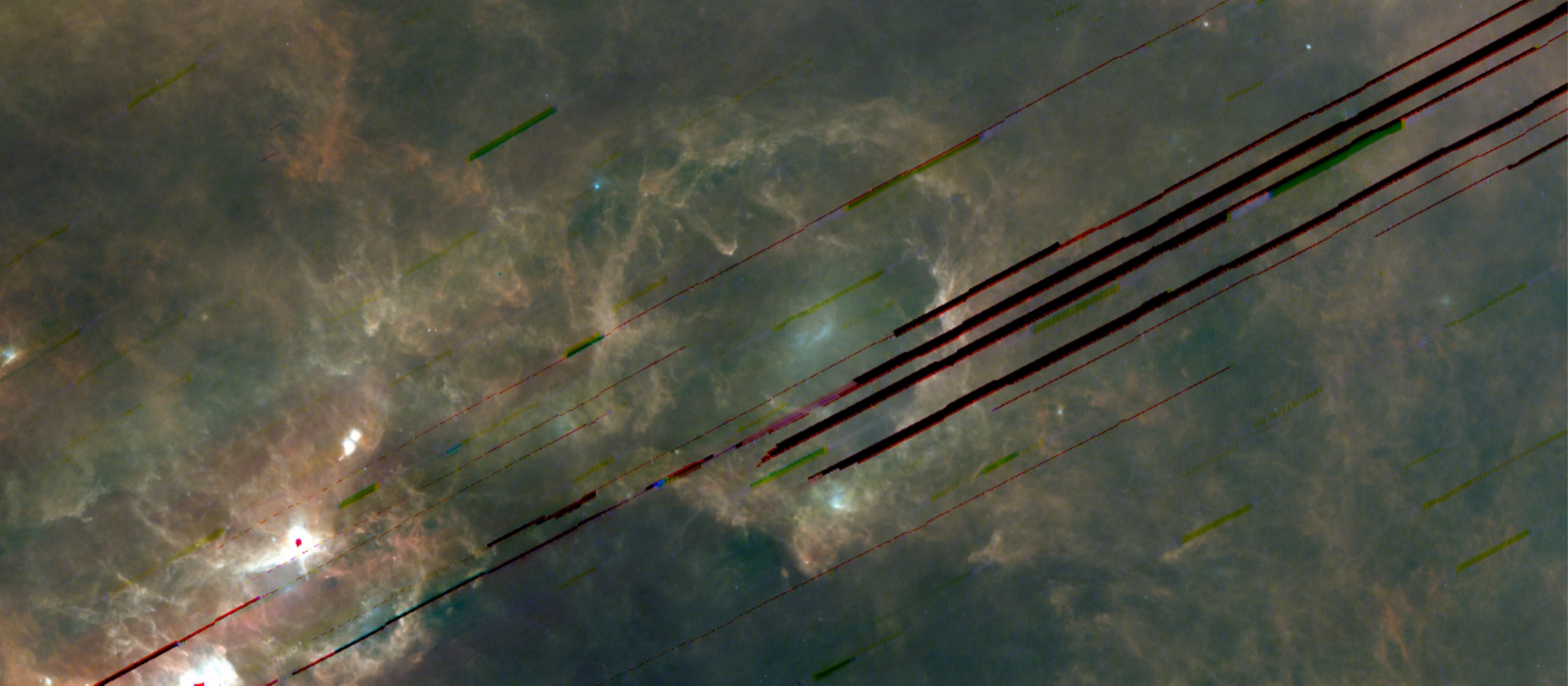}
          \end{center}
          \caption{The $\lambda$~Orionis region and its surroundings in AKARI FIS data, where A65 is blue, A90 is green, and A140 is red. 
          Missing stripe artifacts are visible, and affect all 3 bands shown here as well as the A160 data. }
          \label{fig:LOri_FIS_color}
        \end{figure*}

\begin{figure*}
        \begin{center}
          \includegraphics[width=\textwidth]{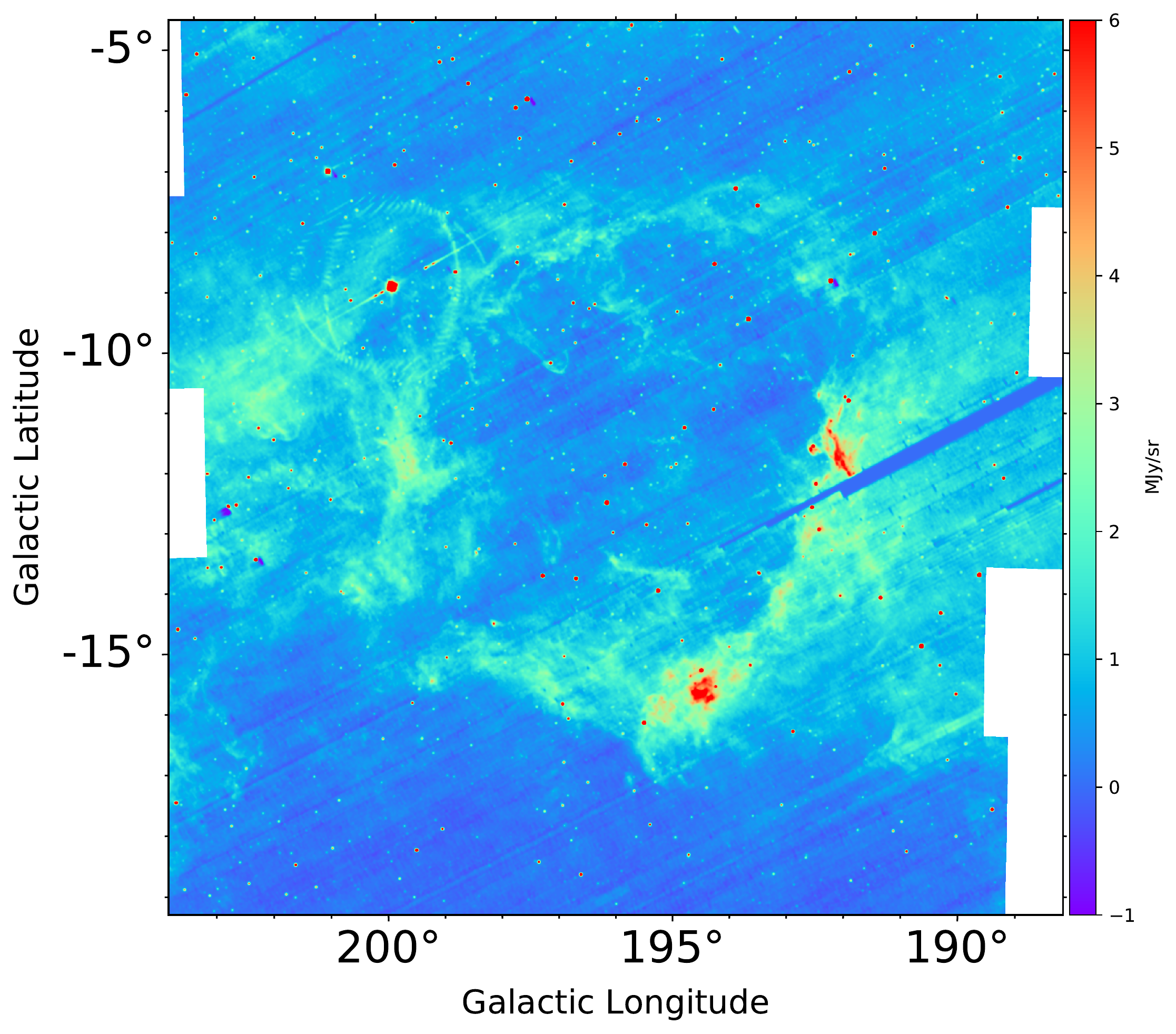}
        \end{center}
          \caption{The $\lambda$~Orionis region in the A9 band at near-native resolution, in galactic coordinates. 
          This is a mosaic created from the 3x3 degree all-sky survey tiles by Ishirara et al. (in prep.) 
          Missing stripes are less of a problem than with the A18 band. 
          The color-scaling is linear.}
          \label{fig:lori_A9_mosaic_smooth}
        \end{figure*}

\begin{figure*}
          \begin{center}
          \includegraphics[width=\textwidth]{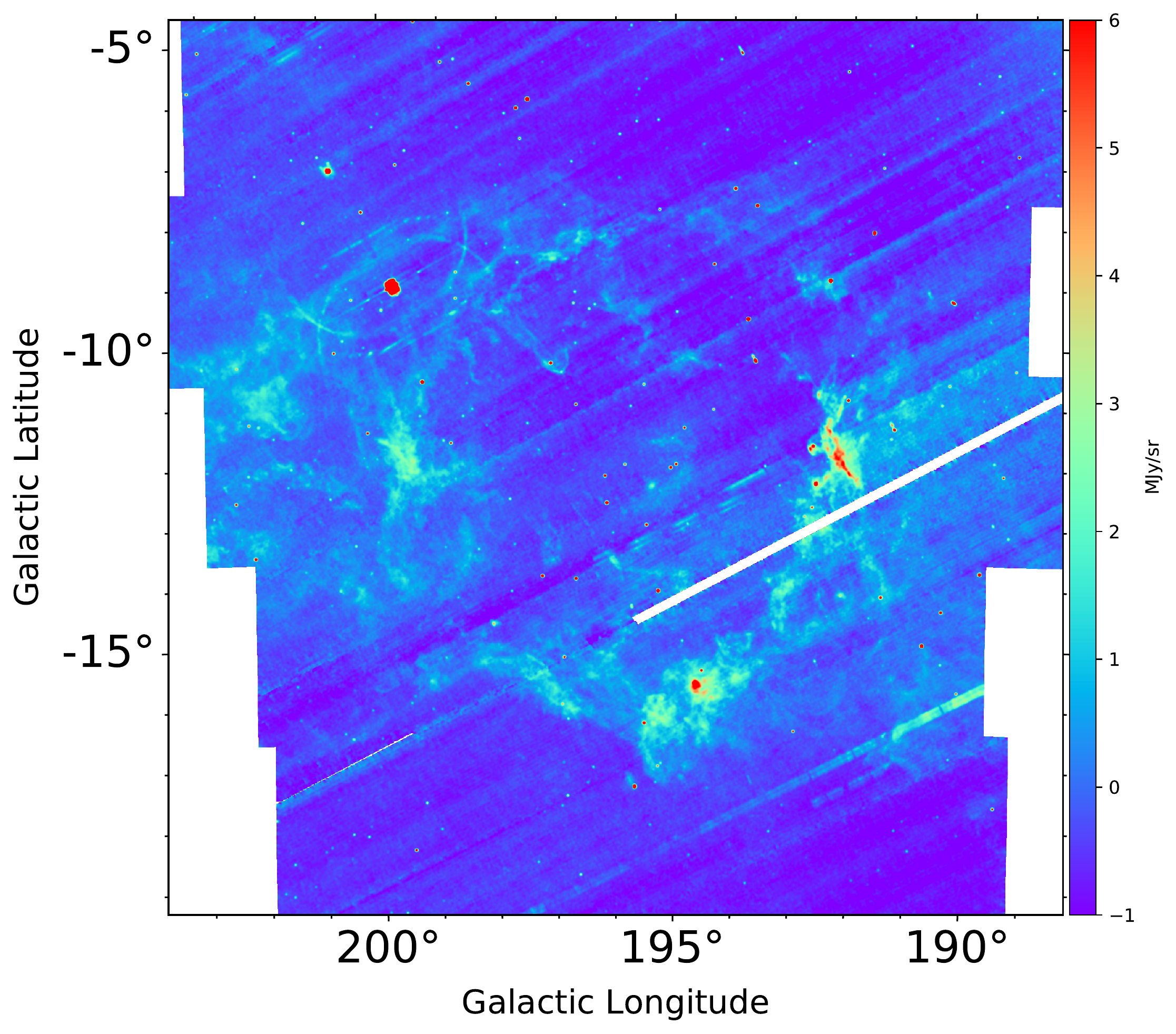}
          \end{center}
          \caption{The $\lambda$~Orionis region in the A18 band at near-native resolution, in galactic coordinates, demonstrating the regions affected by missing stripe errors, but less affected by point-sources than A9 (Fig.~\ref{fig:lori_A9_mosaic_smooth}). 
          The color-scaling is linear.}
          \label{fig:lori_A18_mosaic_smooth}
        \end{figure*}

      The AKARI all-sky survey suffers from a few missing stripe errors throughout the IRC and FIS maps \citep{ishihara10, doi15}. 
      This is a more serious issue for FIS. 
      Unfortunately for the present work, some of these stripes pass directly through the $\lambda$~Orionis region. 
      \Cref{fig:LOri_FIS_color,fig:lori_A9_mosaic_smooth,fig:lori_A18_mosaic_smooth} display the data at near-native resolution, demonstrating where these patterns occur. 
      Additionally there are some saturated pixels in both IRC and FIS data.

      A caveat that comes with added ionized PAH feature coverage of the A9 band, is that the shorter central wavelength placement allows more contamination from point sources. 
      We identify point sources with a moving-window approach provided in the {\tt astropy} Python module \citep{astropy13, astropy18}, flagging pixels which have higher than 5$\sigma$ intensity among the surrounding 100 pixel window. 
      We then place a mask at the center of the flagged point-sources. 
      The masks are propagated through the regridding step, such that the low-resolution pixels having more than 50\% of their area masked in the high resolution tiles, also become masked.
      Such pixels appear red in Fig.~\ref{fig:orionis-akari9}. 
      The same process is applied to the A18 data.

      For the bands extracted from HEALPix files, we first regrid from HEALPix to rectangular grids, and then apply the point-source search and masking as above. 
      For the I12 and I25 images, the rejected pixels were fewer than with A9, but positions overlapped with those already masked in A9. 
      For bands longer than I25, this process did not result in rejected pixels. 
      For the D12 and D25 bands, which are natively at a much lower resolution than AKARI or IRAS, point-sources present more of a challenge. 
      For these images, we visually inspect and mask 3 regions with bright point source contamination consistent with the DIRBE beamsize and with point-sources identified in IRAS and IRC images. 
      Pixel positions masked in any single image, are masked in all of the images before the final analysis.

      We smooth the pixels in the spatial domain, to have a 1$^{\circ}$ FWHM PSF, in order to have a resolution approximating that of the PC AME data.
      The smoothing process relies on the {\tt convolution} module provided in the {\tt astropy} package. 
      We use a simple circular Gaussian kernel for the smoothing process. 
      While asymmetries may be present in the effective beam shapes of the IR bands used, the target resolution of the AME data is large enough relative to the native resolution of the input IR data (especially A9 and A18, see Fig.~\ref{tab:data}) to render the beam shapes and positional variations negligible. 
      Finally, we mask pixels along the edge of the field-of-view where the convolution process produces artifacts.

  \subsection{Background subtraction}
\begin{figure*}
        \begin{center}
          \includegraphics[width=0.94\textwidth,trim={4cm 5cm 3.5cm 5cm},clip]{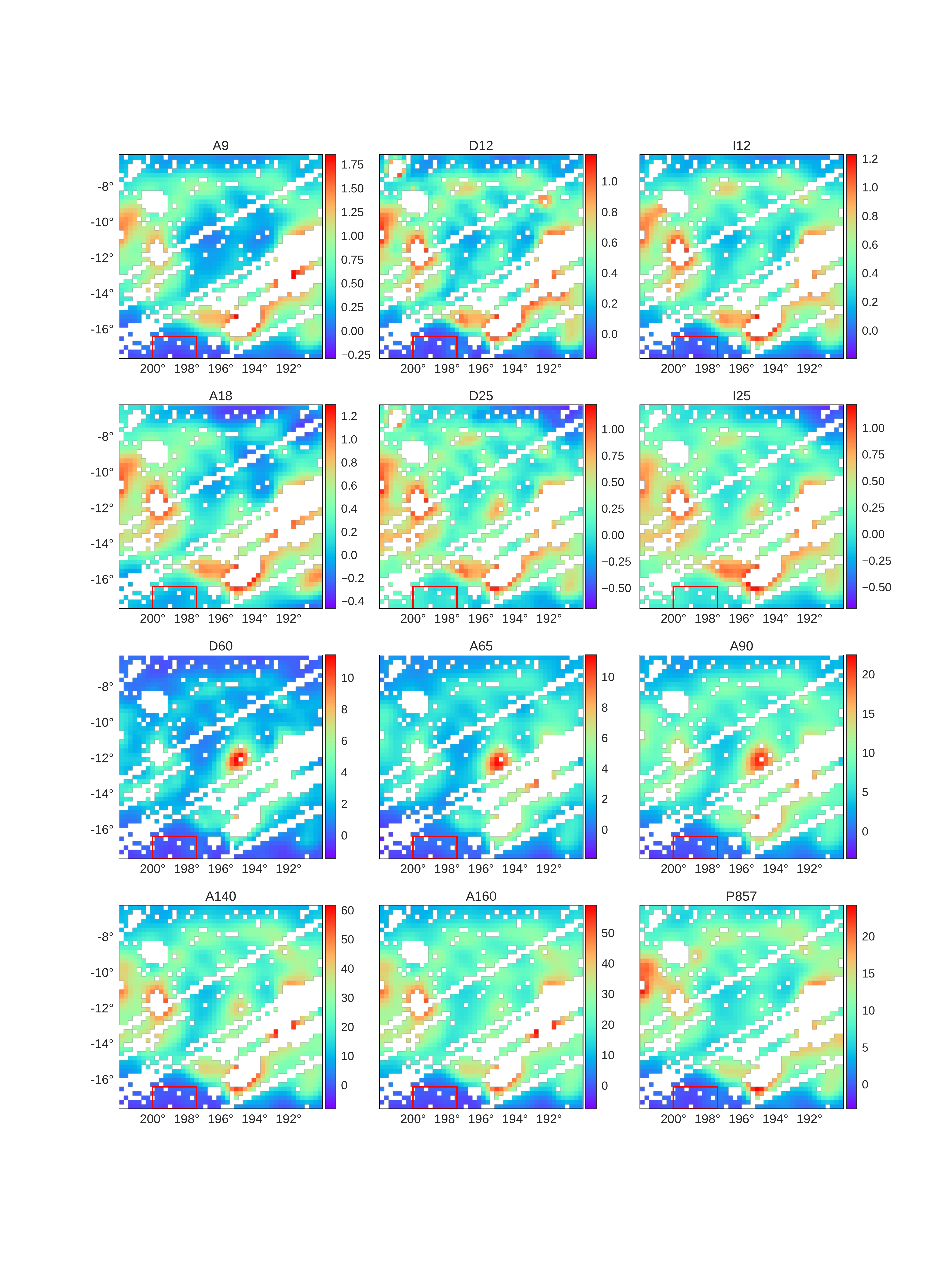}
          \end{center}
          \caption{Processed data at each wavelength for $\lambda$~Orionis. 
          A flat background has been subtracted from each frame based on the mean of pixels within the red rectangle. 
          The pixel width is 0.25$^{\circ}$, with the data PSF smoothed to 1$^{\circ}$ spatial resolution. 
          Masked pixels (point sources, stripe errors, convolution artifacts) are shown as white. The frames share the same FOV and galactic coordinate tangential projection. 
          Colorbars indicate the intensity in MJy/sr, with a linear color-scaling.}
          \label{fig:lori_processed_all}
        \end{figure*}

     In treating background emission, we opted for the simplest approach. We estimate and subtract a mean, flat background level for the region.
     The background level is estimated as the mean of pixels in an `OFF` zone. 
     The final images are shown in Fig.~\ref{fig:lori_processed_all}, with the full mask applied (masked pixels are indicated in white), and with the OFF zone indicated by the red rectangle on each frame.
     Given the spatial resolution and field of view used in our analysis, we judged this simple approach to be the most appropriate.
     More complex background models may be unnecessarily contrived for this analysis.
      We do not expect simple band-by-band intensity correlation tests with the AME to be sensitive to background and foreground emission along the line of sight towards the $\lambda$~Orionis region.
      The results of the dust SED fitting, for a small number of the fainter pixels in our field, may be affected by this background subtraction. 
      This applies to the absolute values returned by the SED fitting, however we are primarily interested in the relative variations in correlation strengths between the AME and dust parameters.
      Since most of the pixels however are bright relative to this background level, we do not expect the background subtraction to affect the correlation analyses applied to the SED fitting results described in Sec.~\ref{sec:hb_analysis}.
      The general morphology as seen in the high resolution AKARI data (\Cref{fig:LOri_FIS_color,fig:lori_A9_mosaic_smooth,fig:lori_A18_mosaic_smooth}) remains well pronounced in the final, low resolution images.

\section{Analysis}
\label{sec:analysis}
  For the present work, we consider the spinning PAH hypothesis to have the highest degree of testability, due to the well-established presence of aromatic emission features in the ISM. 
  We do not argue against the physical plausibility of nanosilicates to produce the AME. 
  Indeed, there is no argument to date that these potential physicalities are mutually exclusive, as long as both potential carriers are sufficiently abundant. 
  Nor does spinning dust emission theoretically exclude magnetic dipole emission or microwave thermal dust emissivity fluctuations.

\subsection{Investigative approach}
\begin{figure*}
      \begin{center}
        \includegraphics[width=\textwidth]{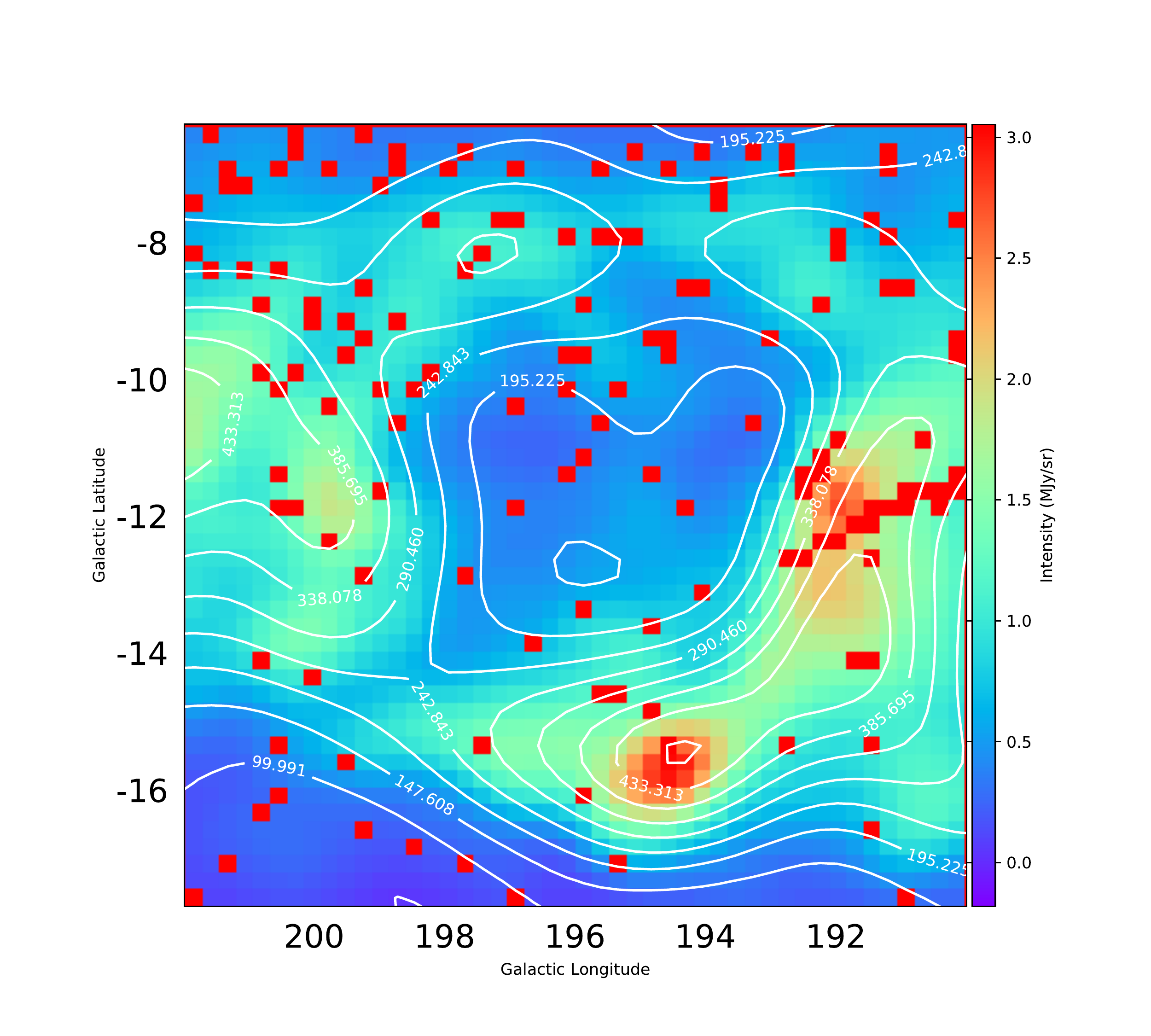}
        \end{center}
        \caption{$\lambda$~Orionis as it appears in the 
        A9 data.
        Contours indicate the AME, as given by the Planck~PR2 AME map.
        The image is smoothed to a 1$^{\circ}$ PSF (much larger than the original 10~arcsec resolution map).
        Contours indicate AME emission, from the variable frequency component, in $\mu{}K_{\rm RJ}$. 
        The color-scaling is linear.
        Masked pixels are indicated in red. Note that the mask applied here differs from the full mask applied in the analysis i.e. the mask applied to the top left panel of Fig.~\ref{fig:lori_processed_all}. Background subtraction has not been applied to this image.
        The $\lambda$~Orionis star itself is approximately located at the center of the image.}
        \label{fig:orionis-akari9}
      \end{figure*}

\begin{figure*}
     \begin{center}
       \includegraphics[width=\textwidth]{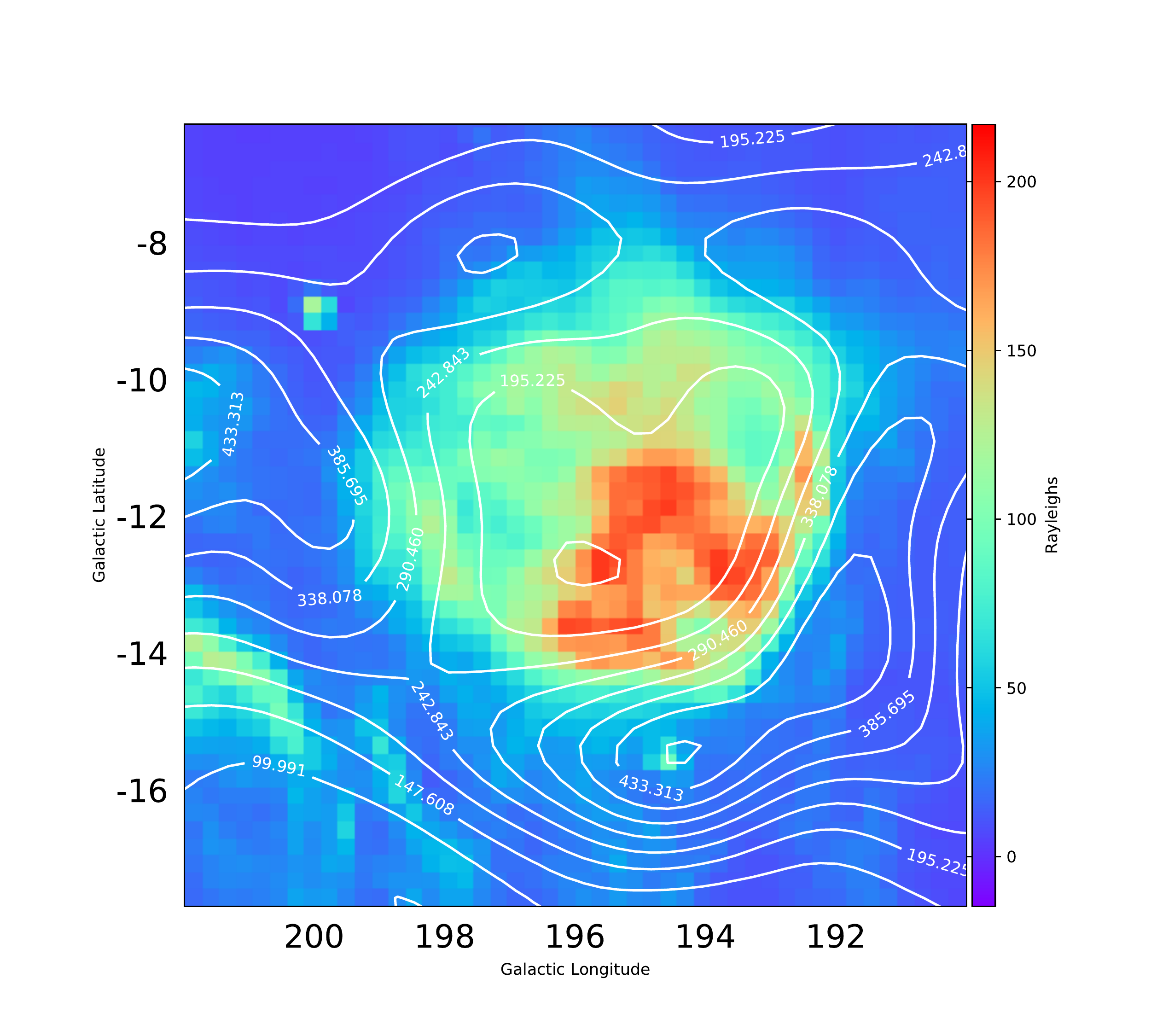}
       \end{center}
       \caption{$\lambda$~Orionis as it appears in $H{\alpha}$ emission by \protect{\citet{finkbeiner03}}. 
       Contours indicate AME emission, from the variable frequency component, in $\mu{}K_{\rm RJ}$. 
       The colorbar indicates $H\alpha$ emission in Rayleighs, with a linear color-scaling.}
       \label{fig:LOri_halpha_AMEvarContours}
     \end{figure*}

  We have carried out an initial comparison of the AME of this region with its 
  MIR to FIR
  dust emission. 
  To demonstrate the structure of the AME relative to the A9 emission, at common spatial resolution, Fig.~\ref{fig:orionis-akari9} as it appears in 1$^{\circ}$-smoothed A9 data. 
   The ring structure itself indicates excess microwave emission attributed to AME, from the dominant variable frequency component $AME_{\rm var}$ (see Sect.~\ref{sec:datasources}). 
   The central region is dominated by free-free emission \citep{aran09, koenig15}. 
   Free-free emission coming from the HII phase surrounding the $\lambda$~Orionis association dominates the region's morphology in Low Frequency Instrument (LFI) images \citep{planck15XXV}. 
   Following a suggestion by \citet{planck15XXV} that this may be among the more reliably component-separated regions, we evaluate if there is any preferential relationship between any parameter of dust emission and the AME. 
   Fig.~\ref{fig:LOri_halpha_AMEvarContours} shows the expected distribution of free-free emission in the region, assuming that $H\alpha$ line emission is a tracer of microwave free-free. 
   This indicates that free-free is strong within the central region, where radiation fields are more intense, and AME is minimal. 
   The strongest AME follows the ring-shaped morphology outside the central bubble of free-free emission.
    We will first show an analysis which does not assume a particular dust IR SED model, comparing only the IR photometric intensities from the 
    MIR to FIR
    to the AME. 
    After comparing all of these intensities, we describe which photometric band shows the best intensity correlation with AME. 
    The core results however will be based on a dust SED modelling analysis, which fits the total abundance of dust, PAH absolute abundance, PAH ionization fraction, and $U$. 
    This is achieved by taking into account all of the photometric intensities, noise and calibration errors, and fitting an SED for all unmasked pixels across the $\lambda$~Orionis region. 
    We will then discuss the core result, suggesting that PAH mass has a stronger relationship with AME than the total dust mass does, representing the first time such a result has been found for a specific AME-prominent region.
    Each analysis method will be outlined in the following subsections, and their results described in \Cref{sec:results}.

	\subsection{Multi-wavelength characterization}
    Our initial exploratory comparison is a simple cross-correlation matrix, based on the Spearman rank correlation coefficient, $r_{s}$ as defined below:
    
\begin{equation}
    r_{s} = \frac {cov(rg_X,rg_Y)} { \sigma_{rg_X} \sigma_{rg_Y} }
\end{equation}
       The ranks of two variables $X$ and $Y$ are indicated by $rg_{Y}$ and $rg_{X}$. $cov$ is the covariance, and $\sigma{}$ the standard deviation. 
       We utilize this test in Python via the {\tt stats.spearmanr}\footnote{https://docs.scipy.org/doc/scipy/reference/generated/scipy.stats.spearmanr.html} function within the {\tt SciPy} library \citep{scipy01}. 
       We assemble a cross-correlation matrix showing $r_s$ for each IR bands vs. each other IR band and the AME.

    \subsection{Bootstrap analysis}
        To assess the robustness of the correlation scores, we employ the Bootstrap re-sampling approach. 
        he ``Bootstrap'' test was first introduced by \citet{efron79}. 
        \citet{feigelson13} gives an updated description of the Bootstrap test within an astronomical context. 
        The process involves creating $N$ random re-samples of a population $P_{0}$, where the size of the re-samples $n_{\rm resamp}$ equals the size $n_{\rm orig}$ of $P_{0}$. This is carried out such that, when a data point is selected from $P_{0}$ and placed in the $Nth$ re-sample $P_{i}$, it immediately becomes eligible for re-selection. 
        This is known as ``resampling with replacement``. 
        Within the re-sampled populations $P_{i}$, some of the original data may be omitted, while others may be over-represented. 
        Ideally this process would be repeated for all of the re-sampling permutations, or $N = nPn$, however this becomes computationally unfeasible with large datasets. 
        $n*log(n)$ resamplings has become a conventional compromise, to sufficiently sample the distribution of a statistic in a reasonable time. 
        Bootstrap tests provide an uncertainty estimate on correlation tests and de-weight outliers.

        We carry out bootstrap correlation tests for each IR band's intensity vs. $AME_{\rm var}$. 
        The data are resampled 10,000 times for each correlation test, a sufficient resampling given the unmasked pixel count of \textasciitilde{}1400 pixels, and considering that the effective beams are oversampled. 
        The $r_{s}$ score calculated for each resample is the same as for the correlation matrix described above. 

        \subsection{Hierarchical Bayesian Dust SED Fitting}
        \label{sec:hb_analysis}
        
           We performed a full pixel by pixel dust SED fitting on the $\lambda$~Orionis photometry, using the HB dust model described in \citet{galliano18a}. 

           For this particular region, we assume a uniformly illuminated dust mixture, {\it ie} the dust mixture in a given pixel is exposed to a single ISRF intensity \citep[see][Sect.~2.4.4]{galliano18a}. 
           We have tried more complex approaches, assuming a distribution of ISRFs inside each pixel \citep[see][Sect.~2.4.5]{galliano18a}. 
           However, this resulted in a narrow ISRF distribution, indicating that a uniformly illuminated ISRF was sufficient.
           
           As for the grain property framework, we assume a mixture of silicate and carbonaceous dust. 
           Instead of the graphite-based carbon dust invoked by the canonical \citet{draine07} model (DL07), we assume the more emissive grain properties of the amorphous carbon (AC) model of \citet{galliano11}. In past years, studies conducted with Herschel Space Observatory and Planck showed that DL07 was likely underestimating the grain emissivity by a factor of about 2. 
           This was first shown in the Large Megellanic Cloud by \cite{galliano11} and later in the Milky Way by \cite{planckIntXXIX16}. 
           Other studies have since confirmed the need to revise the canonical dust emissivity \cite[see review][Sect.~3.1.1]{galliano18b}. 

        The dust model used for this work includes 4 free parameters per pixel: total dust mass $M_{\rm dust}$, fraction of dust mass as PAHs $q_{\rm PAH}$, fraction of ionized PAHs $f_{\rm PAH+}$, and the average strength of the ISRF $U$. 
        For the analysis, we then derive the PAH mass as $M_{\rm PAH} = M_{\rm dust}\times{}q_{\rm PAH}$ and the ionized PAH mass as $M_{\rm PAH+} = M_{\rm PAH}\times{}f_{\rm PAH+}$. 
        We are primarily interested in which of the correlations $M_{\rm dust}$ vs. $I_{\rm AME}$ or $M_{\rm PAH}$ vs. $I_{\rm AME}$ is stronger.

        The fitting of this dust model to the observations is performed using the HB approach described in details in \citet{galliano18a}. 
        The motivation of such an approach is to avoid the numerous biases and noise-induced false correlations encountered by traditional least-squares method (LSM) methods \citep[{\it eg}][]{shetty09,galliano18a}. 
        The first HB dust SED model was presented by \citet{kelly12}, but was limited to modified black body models. 
        Classical (non hierarchical) Bayesian inference consists in sampling the likelihood of the dust parameters of each pixel, in order to estimate their expectation values, uncertainties and correlations. 
        The HB method relies on this approach but deals with two classes of parameters: (i)~the dust parameters of each pixel; and (ii)~a set of {\it hyper-parameters} controlling the distribution of these dust parameters. 
        This global distribution of dust parameters is called the {\it prior}. 
        A LSM or classical Bayesian fit is a single-level model, as it only deals with the dust parameters. 
        In a single-level approach, the parameter's probability distribution of each pixel is independent of the other pixels in the region. 
        Introducing hyperparameters allows one to sample a single large probability distribution for the parameters of all the pixels (the {\it posterior}). 
        This way, the information about the distribution of parameters, among the different pixels, has an impact on the likelihoods of individual pixels. 
        In addition, the HB approach allows us to rigorously treat the various sources of uncertainties. 
        In the present case, we have assumed independent Gaussian noise and correlated calibration errors.

        From a technical point of view, we sample the {\it posterior} distribution with a Markov Chain Monte Carlo (MCMC; Gibbs sampling), of length $10^6$ with a $10^5$ burn-in.
        
\section{Results}
\label{sec:results}
In this section we describe the results of the individual analyses introduced in \Cref{sec:analysis}: the correlation matrix, bootstrapping analysis, and HB dust SED fitting. 
The core results are based on the latter; the former two allow for a comparison between methods of different degrees of complexity.  
We will consider only the dominant component $AME_{\rm var}$ (hereafter, simply $AME$). We found that combining the components does not affect the results here.

    \subsection{$r_{s}$ Correlation Matrix}
    \label{sec:corrmatrix_res}
\begin{figure*}
    \begin{center}
      \includegraphics[width=\textwidth]{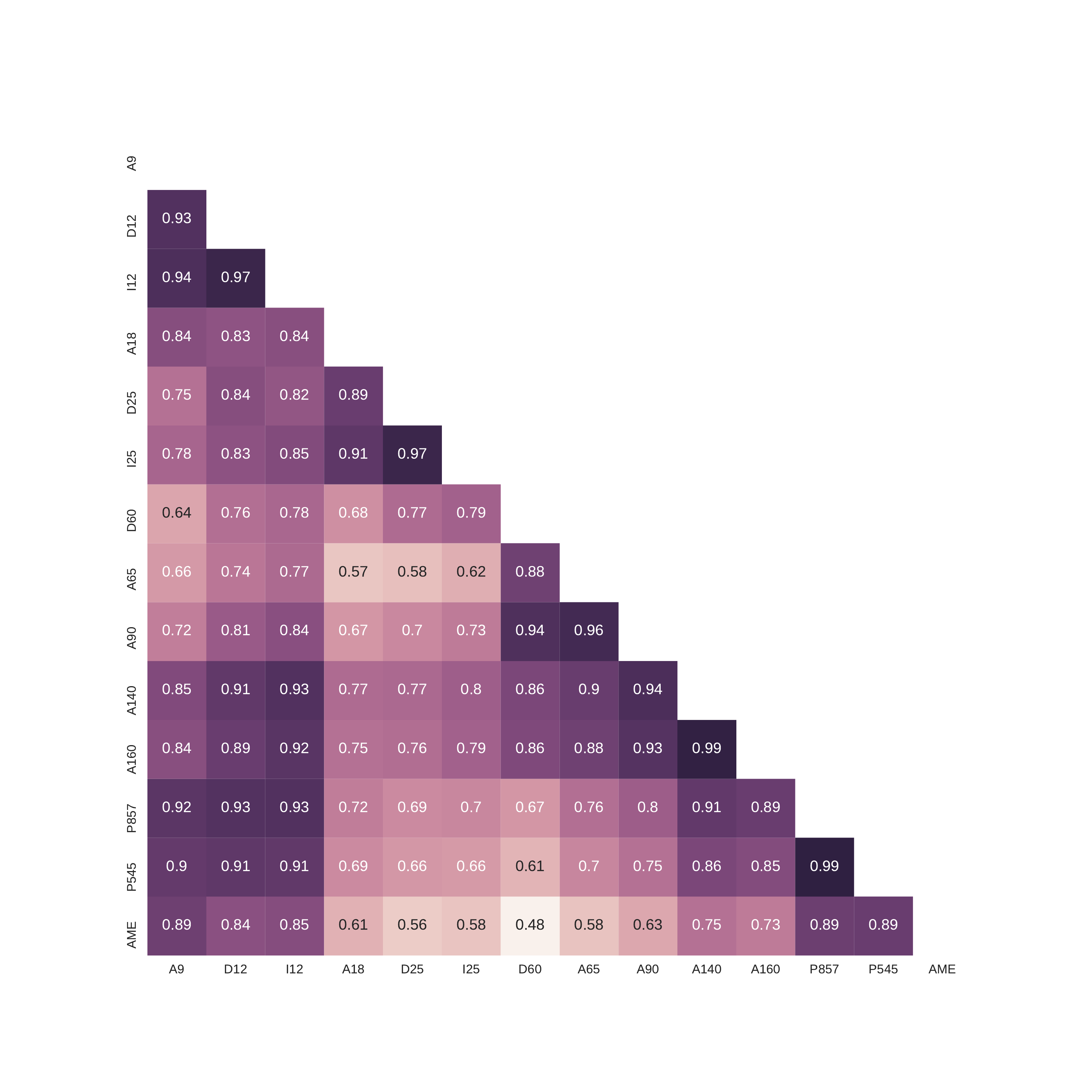}
      \end{center}
      \caption{$r_{s}$ correlation matrix for all of the data used in the $\lambda$~Orionis analysis. 
      The shade and annotation for each cell indicates the $r_{s}$ score. 
      }
      \label{fig:orionis-corr-matrix}
    \end{figure*}

    The correlation matrix corresponding to data shown in Fig.~\ref{fig:lori_processed_all}, are shown in Fig.~\ref{fig:orionis-corr-matrix}. 
    Overall the correlation matrix mirrors the general agreement in spatial morphology between IR and AME. 
    Inspecting the $AME$ row however reveals that the correlation weakens, dropping below 0.80 for the bands between I12 and P857; below 0.6 between I12 and A90.  
    A9, P857, and P545 show the strongest correlations. 
    The overall pattern is that bands dominated by PAH emission (as discussed in Sect.~\ref{sec:datasources}), and those which trace Rayleigh-Jeans thermal dust emission are equally good predictors of the AME. 
    Bands dominated by a mixture of VSGs, and warm dust emission, show a weaker correlation. 
 
    Comparing the images in Fig.~\ref{fig:lori_processed_all}, most of the variation in the correlation scores appears to come from the central region of $\lambda$~Orionis. 
    Because of the known heating present within the ring, from the $\lambda$~Orionis stellar association, and given the brightening of bands between A18 and A90, this variation appears to be due to a temperature increase.

    \subsection{Bootstrapping results}
    \label{sec:bootstrap_res}
\begin{figure*}
        \begin{center}
          \includegraphics[width=\textwidth,trim={3cm 0.25cm 2.5cm 1cm},clip]{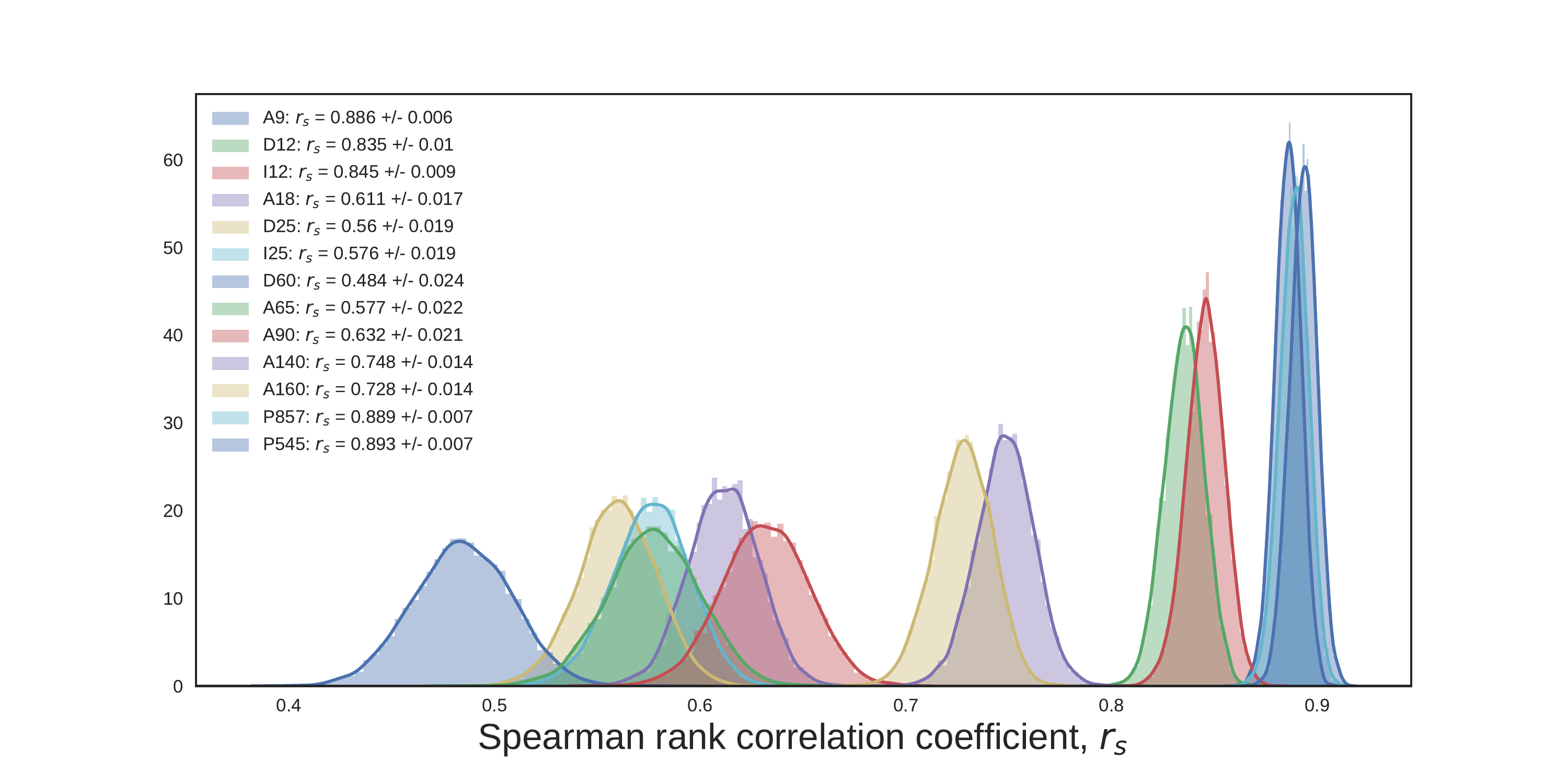}
          \end{center}
          \caption{
          Bootstrap re-sampled correlation tests for IR emission in $\lambda$~Orionis vs. $AME$. 
          Each band's $r_{s}$ distribution is shown in a different color. 
          The width of the distribution indicates the error for the given data in the correlation coefficient. 
          The mean and standard deviation of the scores are given in the legend of each plot.}
          \label{fig:bootstrap_vs_AME}
        \end{figure*}

    The distributions of the bootstrap resamplings are shown in \Cref{fig:bootstrap_vs_AME}. 
    The best correlations are the longest and shortest wavelength bands, consistent with conventional $r_{s}$ scores shown in Fig.~\ref{fig:orionis-corr-matrix}. 
    \Cref{fig:bootstrap_vs_AME} provides not only the centroid $r_{s}$ value for each comparison, but also visualizes the width of the distribution. 
    The strongest two correlations, $r_{s} = {\rm 0.886 \pm 0.006}$ for A9 and $r_{s} = {\rm 0.893 \pm 0.007}$ are overlapping, making a conclusion difficult and justifying the use of the more sophisticated HB dust SED fitting introduced in \Cref{sec:hb_analysis}.

    \subsection{SED Fitting Results}
    \label{sec:hr_res}
\begin{figure*}
              \begin{center}
                \includegraphics[width=\textwidth]{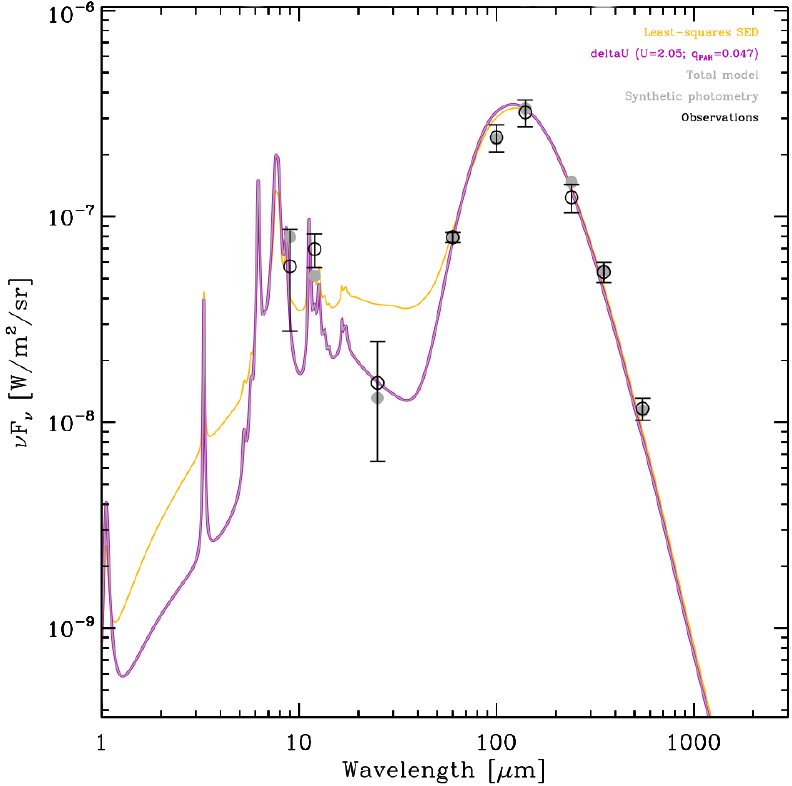}
                \end{center}
                \caption{Observed (black circles and errors) and synthetic photometry (gray filled circles) SED of an example pixel within $\lambda$~Orionis, at {\it l} = 201.8$^\mathrm{o}$ and {\it b} = -16.5$^\mathrm{o}$, along with the dust SED model fit results. Synthetic photometry gives model-predicted values as they would appear if seen through the instrument filters.
                Two SED fits are shown: one for the Bayesian fitting (magenta), and another showing the standard least-squares result for comparison (yellow). 
                The fitted ISRF strength $U$, and fraction of mass in PAHs, $q_{\rm PAH}$ are also given. The}
                \label{fig:fred_LOri_notes_Oct2017_fig1a}
              \end{figure*}

\begin{figure*}
              \begin{center}
                \includegraphics[width=\textwidth]{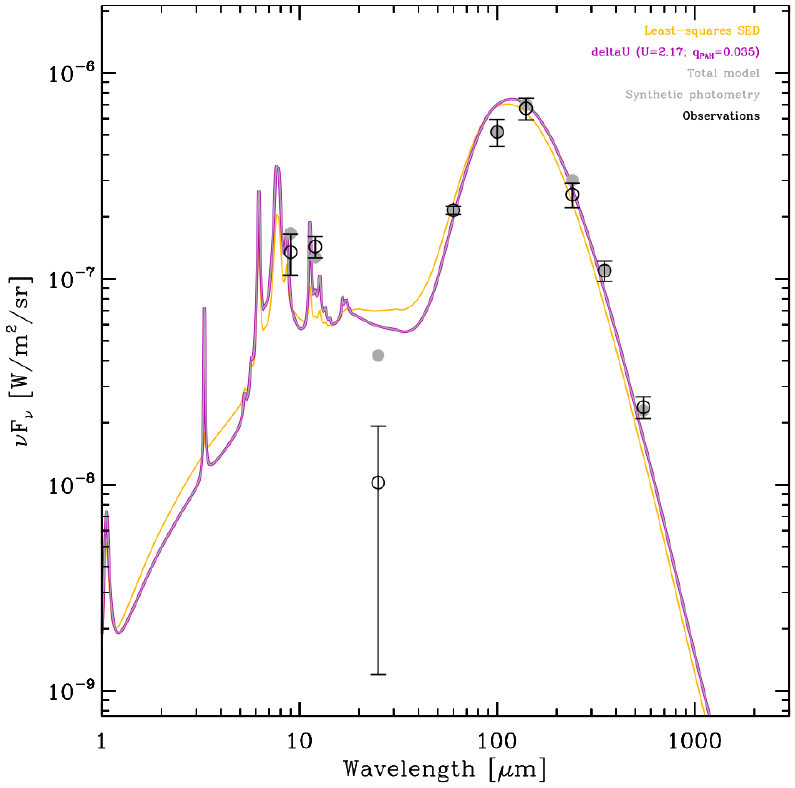}
                \end{center}
                \caption{The same as Fig.~\ref{fig:fred_LOri_notes_Oct2017_fig1a}, but for the pixel at {\it l} = 196.13$^\mathrm{o}$ and {\it b} = -11.90$^\mathrm{o}$, at the center of the $\lambda$~Orionis region.}
                \label{fig:fred_LOri_notes_Oct2017_fig1b}
              \end{figure*}

\begin{figure*}
                \begin{center}
                \includegraphics[width=\textwidth]{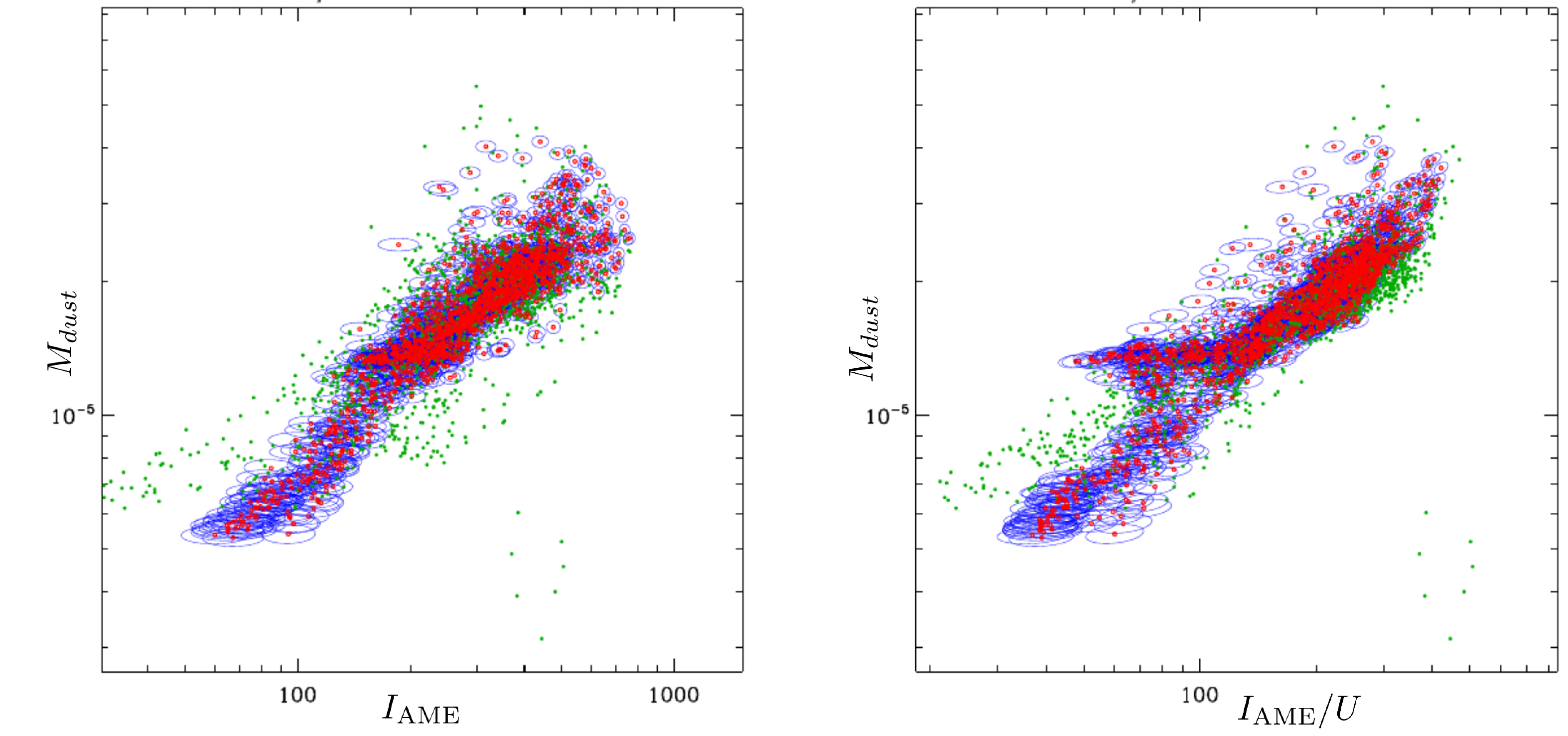}
                 \end{center}
                 \caption{Scatter plot with the bivariate error ellipses generated through the HB SED fitting, of total dust mass $M_{\rm dust}$ vs. $I_{\rm AME}$ normalized by $U$. 
                 Green dots indicate the results when using a simple least-squares method fit, for comparison with the HB method. 
                 $r_{\rm p} = {\rm 0.835 \pm 0.002}$ for the $I_{\rm AME}$ case (left) and ${\rm 0.876 \pm 0.002}$ for the $I_{\rm AME}/U$ case (right).}
                 \label{fig:LOri_SEDfit_Mdust}
               \end{figure*}

\begin{figure*}
              \begin{center}
                \includegraphics[width=\textwidth,trim={0cm 0cm 0cm 0cm},clip]{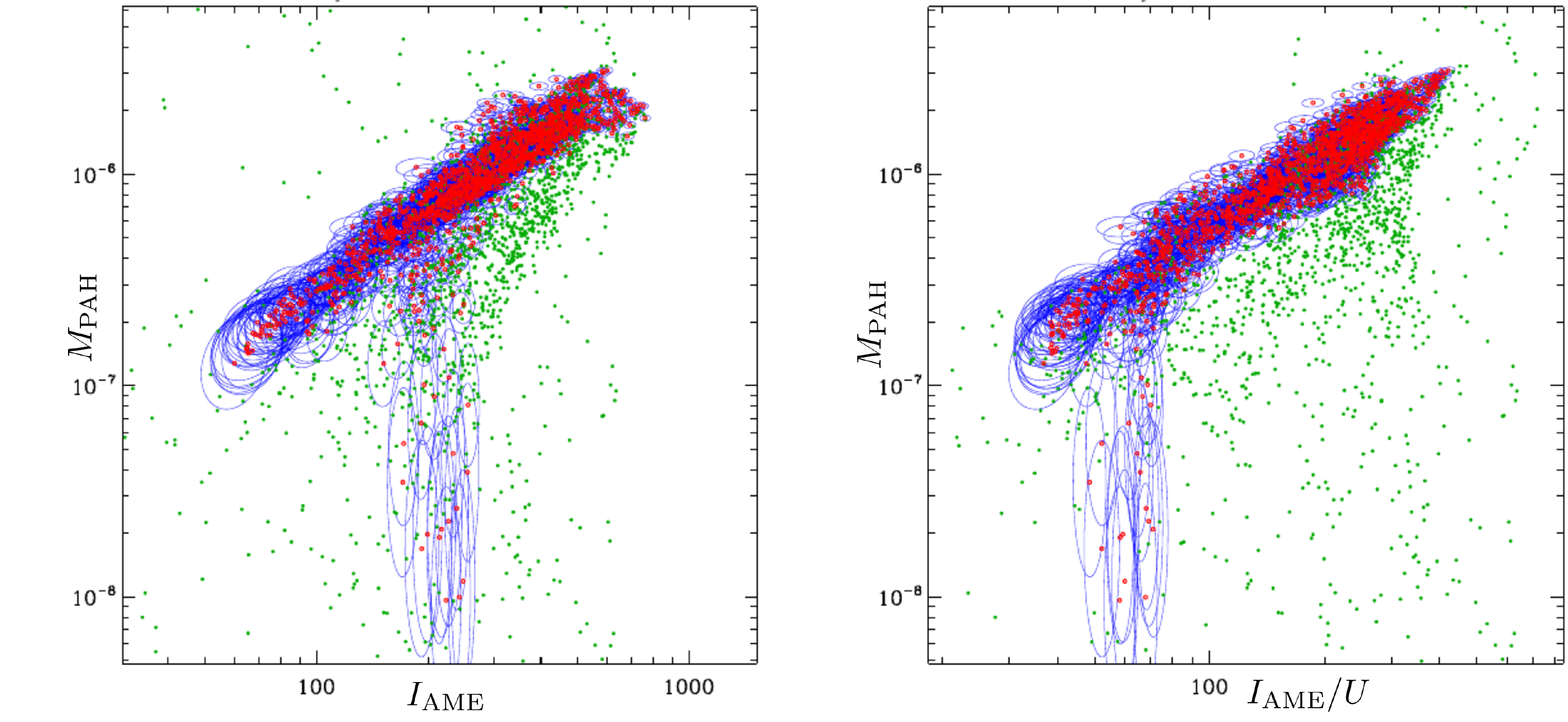}
                \end{center}
                \caption{The same comparison as given by Fig.~\ref{fig:LOri_SEDfit_Mdust}, but showing total mass of PAHs ($M_{\rm PAH}$) rather than total dust mass on the y-axis. 
                $r_{\rm p} = {\rm 0.871 \pm 0.003}$ for the $I_{\rm AME}$ case (left) and ${\rm 0908 \pm 0.003}$ for the $I_{\rm AME}/U$ case (right).}
                \label{fig:LOri_SEDfit_Mpah}
              \end{figure*}

\begin{figure*}
              \begin{center}
                \includegraphics[width=\textwidth]{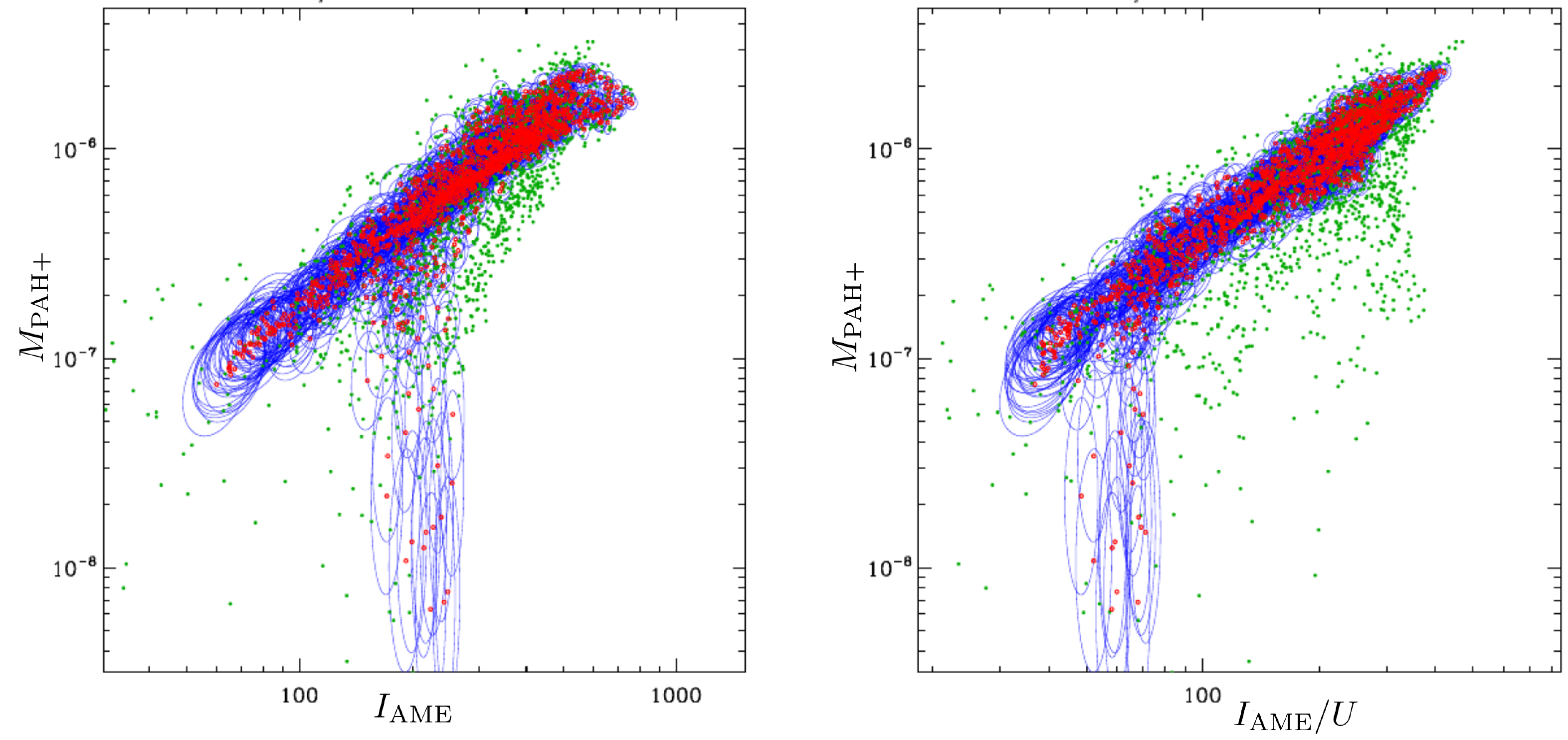}
                \end{center}
                \caption{The same as in Figs.~\ref{fig:LOri_SEDfit_Mdust} and~\ref{fig:LOri_SEDfit_Mpah}, but specifically comparing an estimate of the charged component of PAH mass $M_{\rm PAH+}$. 
                This includes anions and cations, since we cannot distinguish between these two spectroscopically. 
                $r_{\rm p} = {\rm 0.876 \pm 0.006}$ for the $I_{\rm AME}$ case (left) and ${\rm 0915 \pm 0.005}$ for the $I_{\rm AME}/U$ case (right).}
                \label{fig:LOri_SEDfit_Mpahi}
              \end{figure*}

    Two sample SED fitting results are shown in \Cref{fig:fred_LOri_notes_Oct2017_fig1a,fig:fred_LOri_notes_Oct2017_fig1b}. 
    Fig.~\ref{fig:fred_LOri_notes_Oct2017_fig1b} represents a pixel at the center of the region ({\it l} = 196.13$^\mathrm{o}$, {\it b} = -11.90$^\mathrm{o}$). 
    The SED in Fig.~\ref{fig:fred_LOri_notes_Oct2017_fig1b} comes from a pixel outside of the ring ({\it l} = 201.80$^\mathrm{o}$, {\it b} = -16.50$^\mathrm{o}$).
           Performing such fits for all of the pixels, we are able to see how $I_{\rm AME}$ varies with the dust properties of the region.
           Figs.~\ref{fig:LOri_SEDfit_Mdust},~\ref{fig:LOri_SEDfit_Mpah},~\ref{fig:LOri_SEDfit_Mpahi} show how the fitted parameters vary with the AME in each pixel in our analysis. 
           It is important to keep in mind when interpreting the scatter plots in Figs.~\ref{fig:LOri_SEDfit_Mdust},~\ref{fig:LOri_SEDfit_Mpah},~\ref{fig:LOri_SEDfit_Mpahi} that our data is oversampled. 
           As described in \ref{sec:dataproc}, the number of data points pixels in the analysis is greater than the spatial-resolution-limited number of independent measurements in our field of view. 
           Fig.~\ref{fig:LOri_SEDfit_Mdust} shows the fitted dust mass per pixel, relative to the AME intensity. 
           We use a standard Pearson correlation coefficient test, $r_{p}$ to determine the correlation strength between $I_{\rm AME}$ and 3 dust properties of interest $M_{\rm dust}$, $M_{\rm PAH}$, $M_{\rm PAH+}$. 
           We find the values of these scores, hereafter $r_{\rm p, dust}$, $r_{\rm p, PAH}$, and $r_{\rm p, PAH+}$, to be 0.835, 0.871, and 0.876.
            We performed also a comparison in which the AME intensity is normalized by $U$: this increases $r_{\rm p}$ for all 3 comparisons, however the ranking of the correlation scores does not change. 
            For the $U$ normalized case, we find $r_{\rm p, dust}$, $r_{\rm p, PAH}$, and $r_{\rm p, PAH+}$ to be 0.876, 0.908, and 0.915. 
            Although spinning dust emission is not predicted to vary directly with $U$, we consider that $U$ may serve as a diagnostic of environmental conditions in the ISM. Figs.~\ref{fig:LOri_SEDfit_Mpah} and~\ref{fig:LOri_SEDfit_Mpahi} describe the variation with $M_{\rm PAH}$ and $M_{\rm PAH+}$.
    Based on the dust properties derived from these SED fits, we investigate whether any fitted parameter shows a preferential relation with the AME. Figs.~\ref{fig:LOri_SEDfit_Mdust}-\ref{fig:LOri_SEDfit_Mpahi} reveal a very similar trend between the AME and the parameters $M_{\rm PAH}$, $M_{\rm PAH+}$, and $M_{\rm dust}$. 
    There is a slightly improved correlation between $M_{\rm dust}$ and $M_{\rm PAH}$ (0.835 vs 0.871). 
    This is consistent with the intensity cross correlations in Fig.~\ref{fig:orionis-corr-matrix}. 
    These correlations are discussed further in Sect.~\ref{sec:discussion} and Fig.~\ref{fig:LOri_SEDfit_PDFs}.

    \subsection{Comparing the Correlation Strengths}
\begin{figure*}
        \begin{center}
          \includegraphics[width=0.49\textwidth,trim={0.75cm 1cm 0.75cm 1cm},clip]{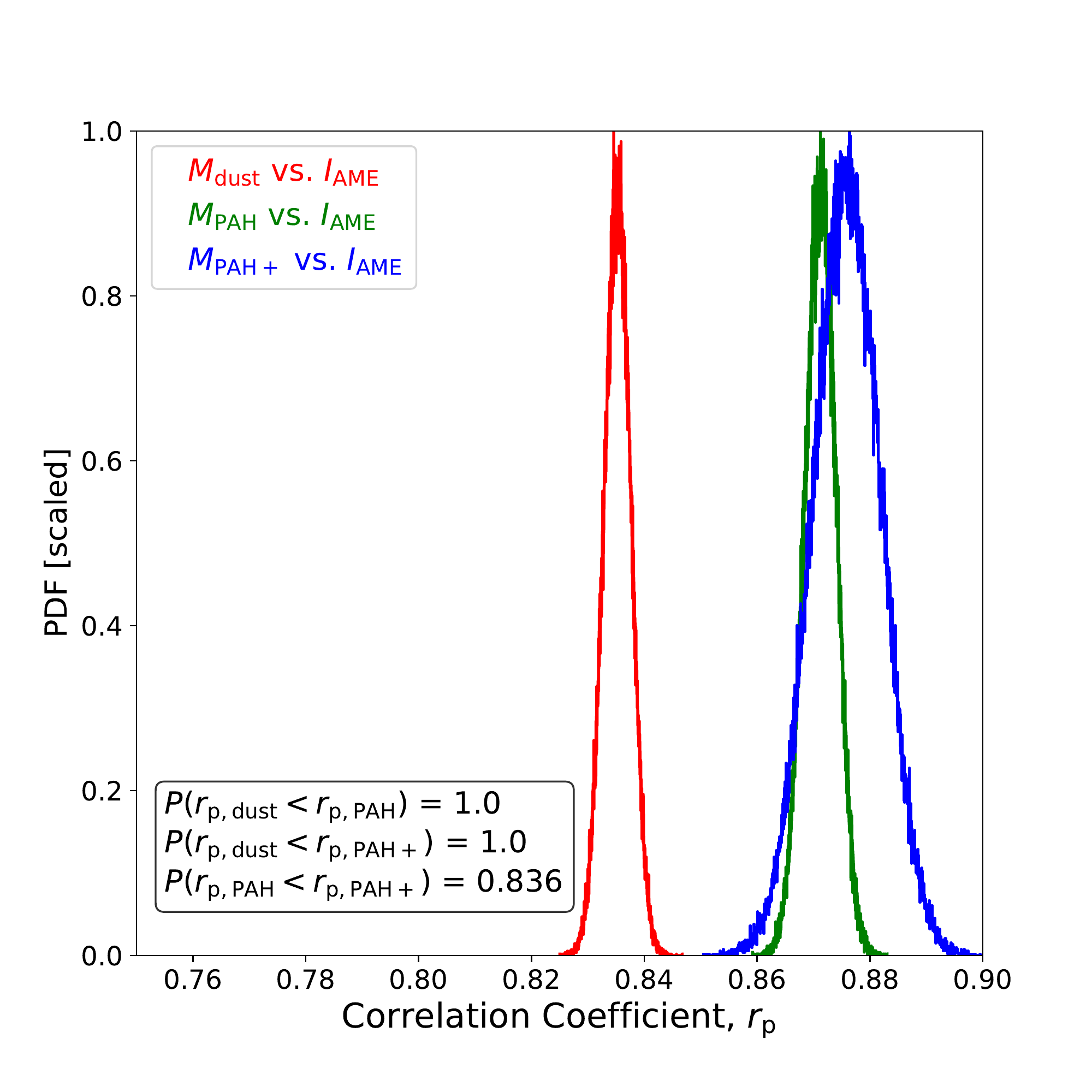}
           \includegraphics[width=0.49\textwidth,trim={0.75cm 1cm 0.75cm 1cm},clip]{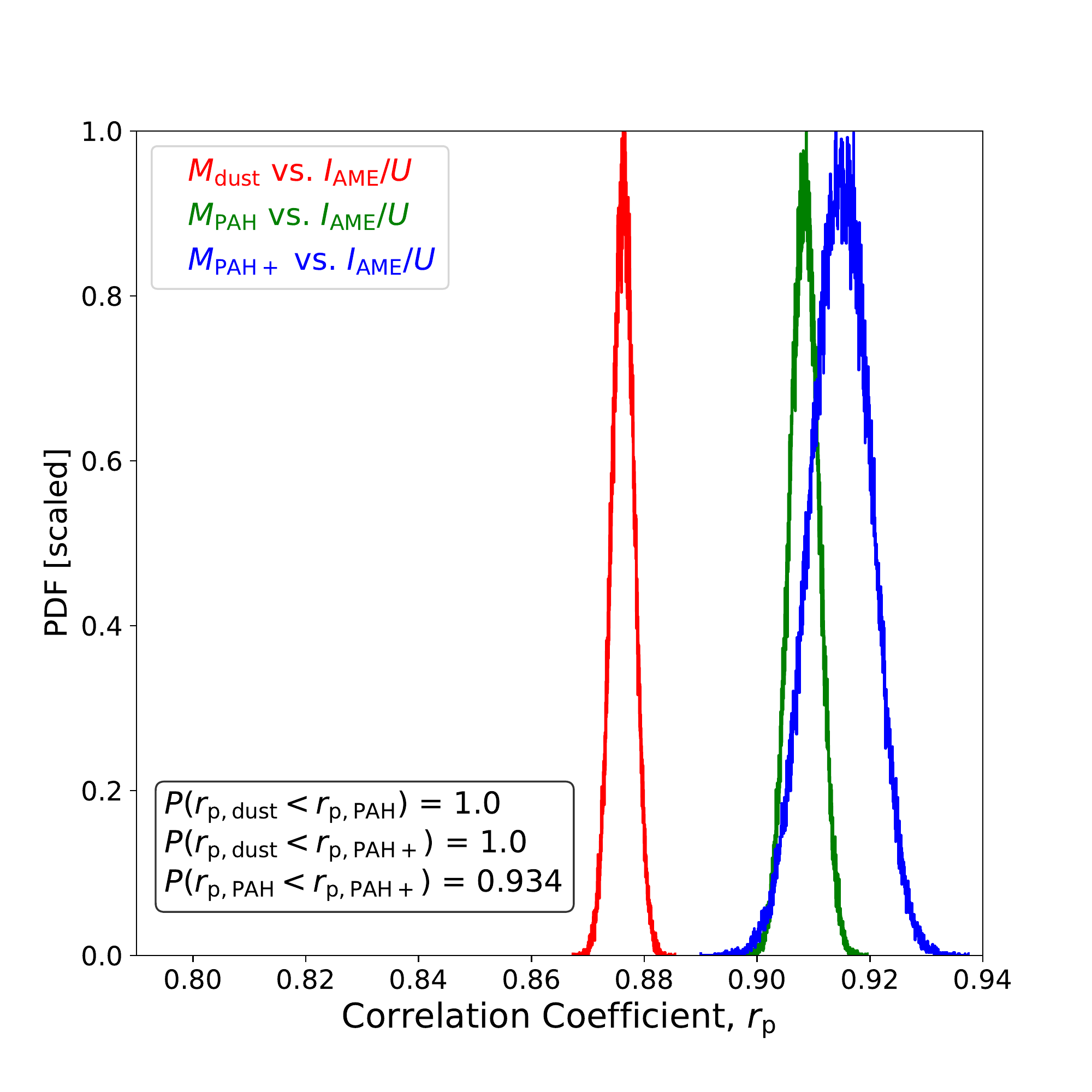}
          \end{center}
          \caption{The HB correlation probability distributions of $r_{\rm p}$) for the three physical parameters vs. the AME intensity: total dust mass, $r_{\rm p, dust}$ (red); total PAH mass $r_{\rm p, PAH}$ (green); and only the ionized PAH mass $r_{\rm p, PAH+}$ (blue). 
          Also given are the probabilities of either PAH component being better correlated with AME than dust mass, as well as the probability that ionized PAH mass correlates better than total PAH.}
          \label{fig:LOri_SEDfit_PDFs}
        \end{figure*}

\begin{figure*}
        \begin{center}
          \includegraphics[width=\textwidth]{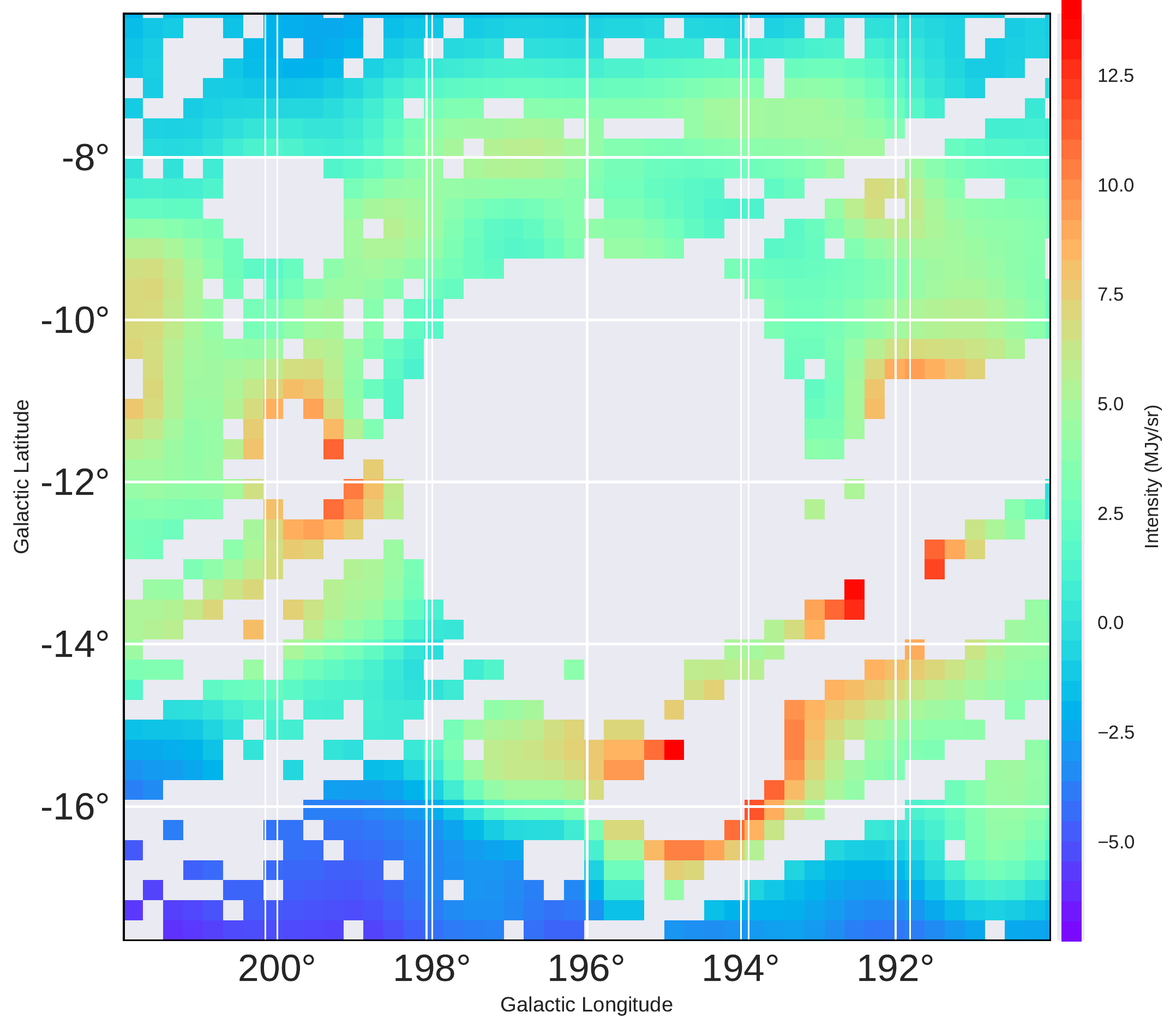}
          \end{center}
          \caption{ The A90 image of $\lambda$~Orionis shown to demonstrate the mask applied to the central region, in order to produce the PDFs in Fig.~\ref{fig:PDFs_center_masked_Iame}. 
          The color-scaling is linear. 
          The region showing an apparent dust temperature peak, which was most prominent in the A90 image and the major point of variation across all the wavelengths, is thus omitted.}
          \label{fig:lambda_orionis_center_mask}
        \end{figure*}

\begin{figure*}
        \begin{center}
            \includegraphics[width=0.49\textwidth,trim={0.75cm 1cm 0.75cm 1cm},clip]{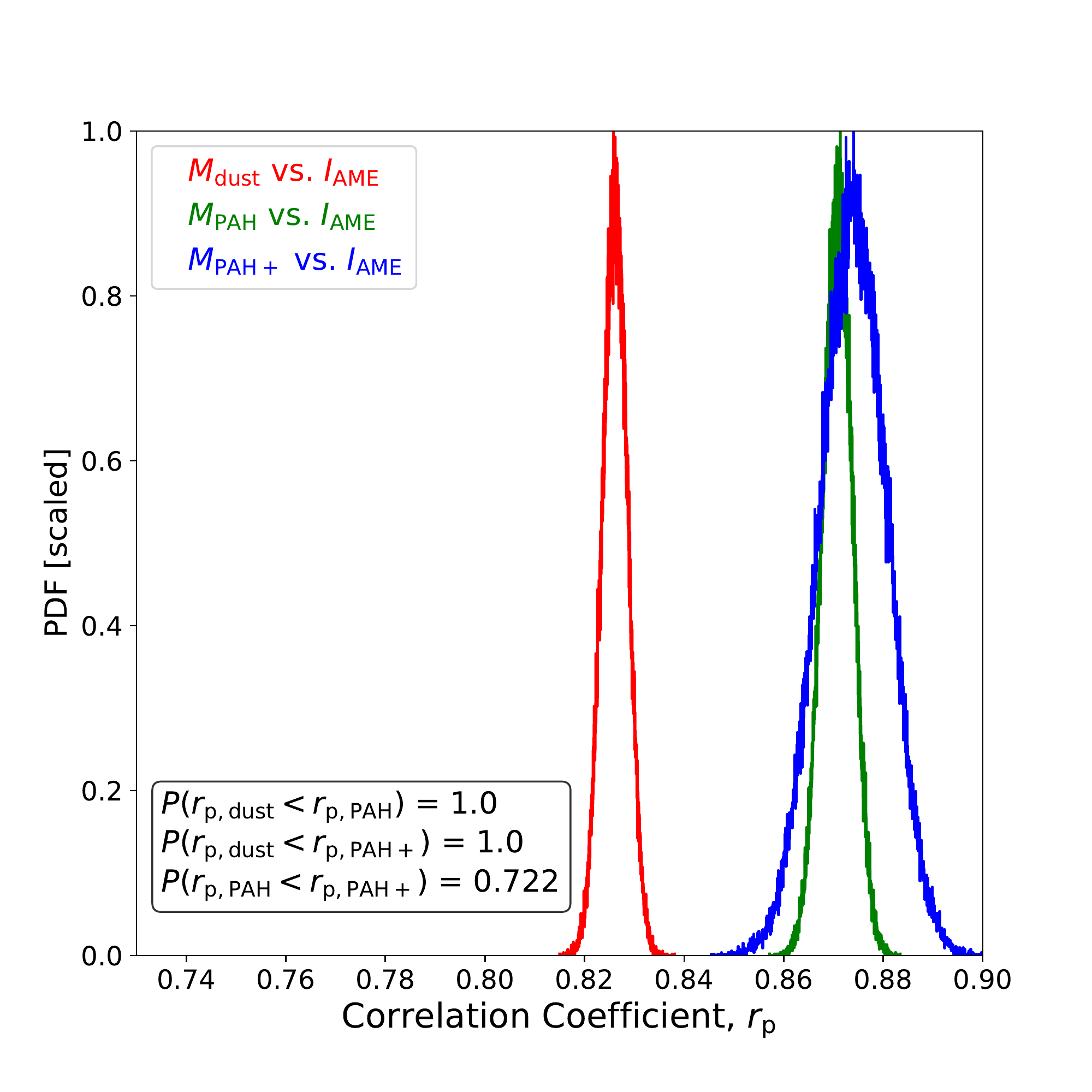}
             \includegraphics[width=0.49\textwidth,trim={0.75cm 1cm 0.75cm 1cm},clip]{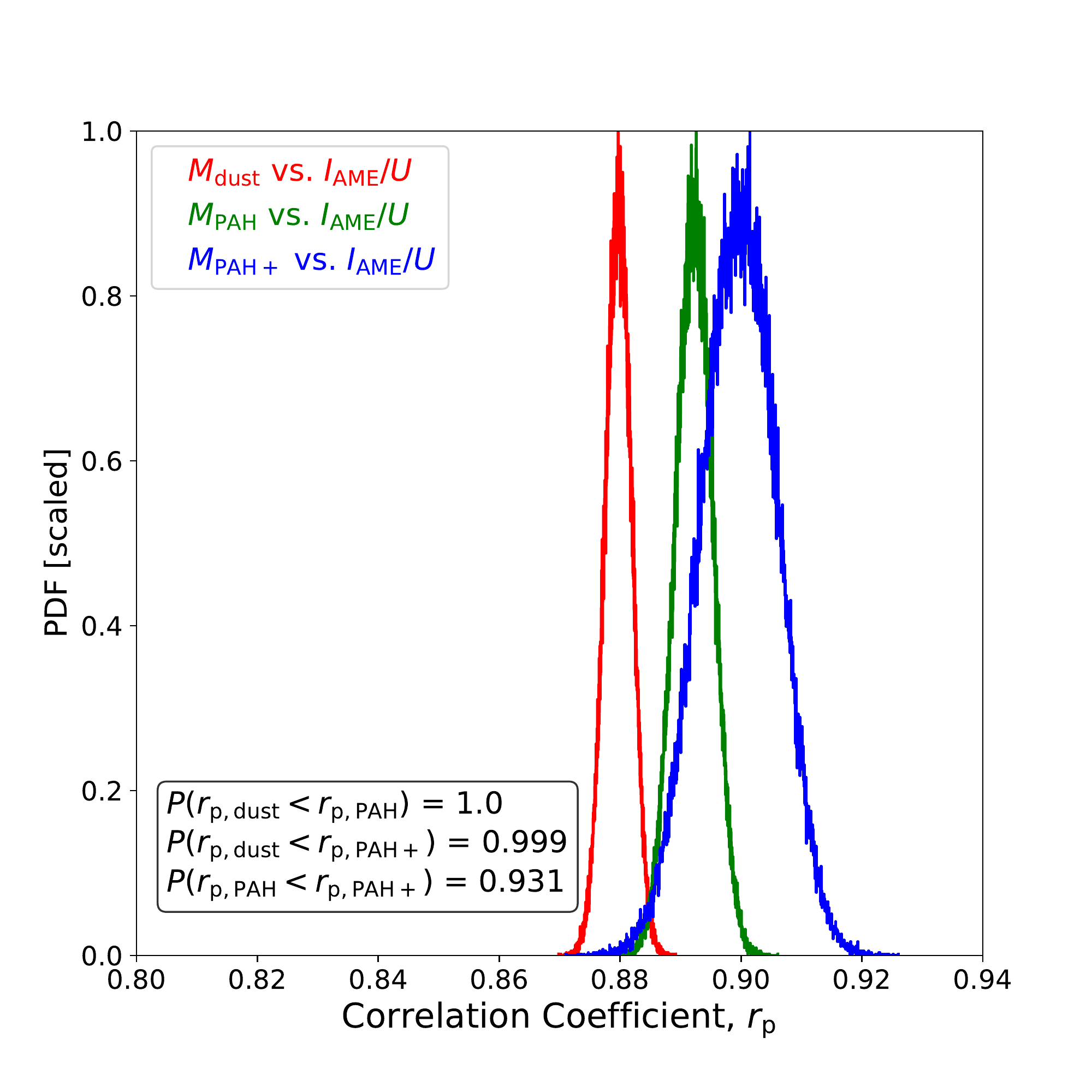}
         \end{center}
         \caption{ The HB correlation coefficent ($r_{\rm p}$) as shown in PDFs as shown in Fig.~\ref{fig:LOri_SEDfit_PDFs}, but with the central region of $\lambda$~Orionis masked, according to Fig.~\ref{fig:lambda_orionis_center_mask}. }
         \label{fig:PDFs_center_masked_Iame}
       \end{figure*}

    In addition to generating a PDF for each of the fitted parameters, HB allows us to plot the final PDF of the correlation coefficients themselves. 
    This is a more robust way of comparing the correlation strengths than simply bootstrapping the correlations. 
    Fig.~\ref{fig:LOri_SEDfit_PDFs} shows PDFs for $r_{\rm p},M_{\rm dust}$, $r_{\rm p},M_{\rm PAH}$, and $r_{\rm p},M_{\rm PAH+}$.
    To assess the effect of the morphology of $\lambda$~Orionis on these results, we also generate PDFs which have the central portion of $\lambda$~Orionis omitted. 
    The extent of the omitted region is shown in Fig.~\ref{fig:lambda_orionis_center_mask}, where the masked pixels are indicated in white.
        The SED fitting MCMC results from these masked pixels are excluded from the PDFs shown in Fig.~\ref{fig:PDFs_center_masked_Iame}.

  \section{Discussion}
  \label{sec:discussion}
      In  $\lambda$~Orionis we found that across the whole region, A9 emission and P545 emission were the most strongly correlated with AME. 
      This is apparent both in the photometric band analysis, and in the dust SED fitting.  
      The fact that the correlation strengths of PAH-tracing mission and sub-mm emission are similar is in-line with what we have seen in \citet{ysard10b} and \citet{hensley16}. 
      In those works, the two relationships (MIR vs. AME and FIR vs. AME) are very close, although these two papers are odds as to which relationship is stronger, and thus in their final interpretation. 
      With the present data and analysis of $\lambda$~Orionis, we fail to rule out PAHs as carriers of the AME. 
      In fact, our primary result, shown in Fig.~\ref{fig:LOri_SEDfit_PDFs} indicates that PAH mass is a stronger predictor of the AME, compared to the total dust mass. 
      This is the first time such a significant result has been found, in support of the PAH hypothesis for a particular region. 
      This core result is unchanged even when we examine the outer ring structure separately from the central region, as shown in Fig.~\ref{fig:PDFs_center_masked_Iame}.
      It is true that the correlation is stronger when the central region, however this may not be surprising given that ring traces the PDR. 
      Such regions are well known to include both PAHs and BGs \citep{allamandola89, bakes94}.

      The results are consistent with a scenario in which PAH mass, cold dust, and the AME are tightly correlated in $\lambda$~Orionis.
      A correlation between cold dust and PAHs is observed in extragalactic targets \citep{haas02}, and may be inferred from the correlation between UIB and FIR emission reported in diffuse galactic ISM \citep{onaka96}.

      In the case that AME emanates from spinning PAHs, it is not surprising that cold dust would also correlate with the AME. 
      Weaker correlation from 25 to 70~$\mu$m may indicate that AME is weaker in regions of warmer dust and stronger radiation fields. 
      Such an anti-correlation with harsher radiation are consistent with the carriers of AME being destroyed in the central region of $\lambda$~Orionis, thus leading to substantially decreased spinning dust emission.
     
     While our result does appear to support the AME-from-PAHs hypothesis in $\lambda$~Orionis, we stress that the relationship between AME and PAH emission should be investigated on a region-by-region basis. 
     Future works should examine the AME of $\lambda$~Orionis at higher resolution, to ensure that the improved correlation (relative to total dust mass) is not seen only at lower resolution, as was found in RCW~175 by \cite{tibbs12b}. 
     Also higher resolution studies could provide a better understanding of the exact contribution from free-free emission in the center of $\lambda$~Orionis.
     Revisiting past analyses of AME regions such as RCW~175 \citep{tibbs12b}, Perseus, or $\rho$~Ophiucus by applying the HB dust fitting approach presented here may help to understand if conventional dust SED fitting methods have overlooked subtle relationships between AME and PAHs.
        
      \subsection{Performance of the Hierarchical Bayesian Fitting}
\begin{figure*}
        \begin{center}
            \includegraphics[width=\textwidth]{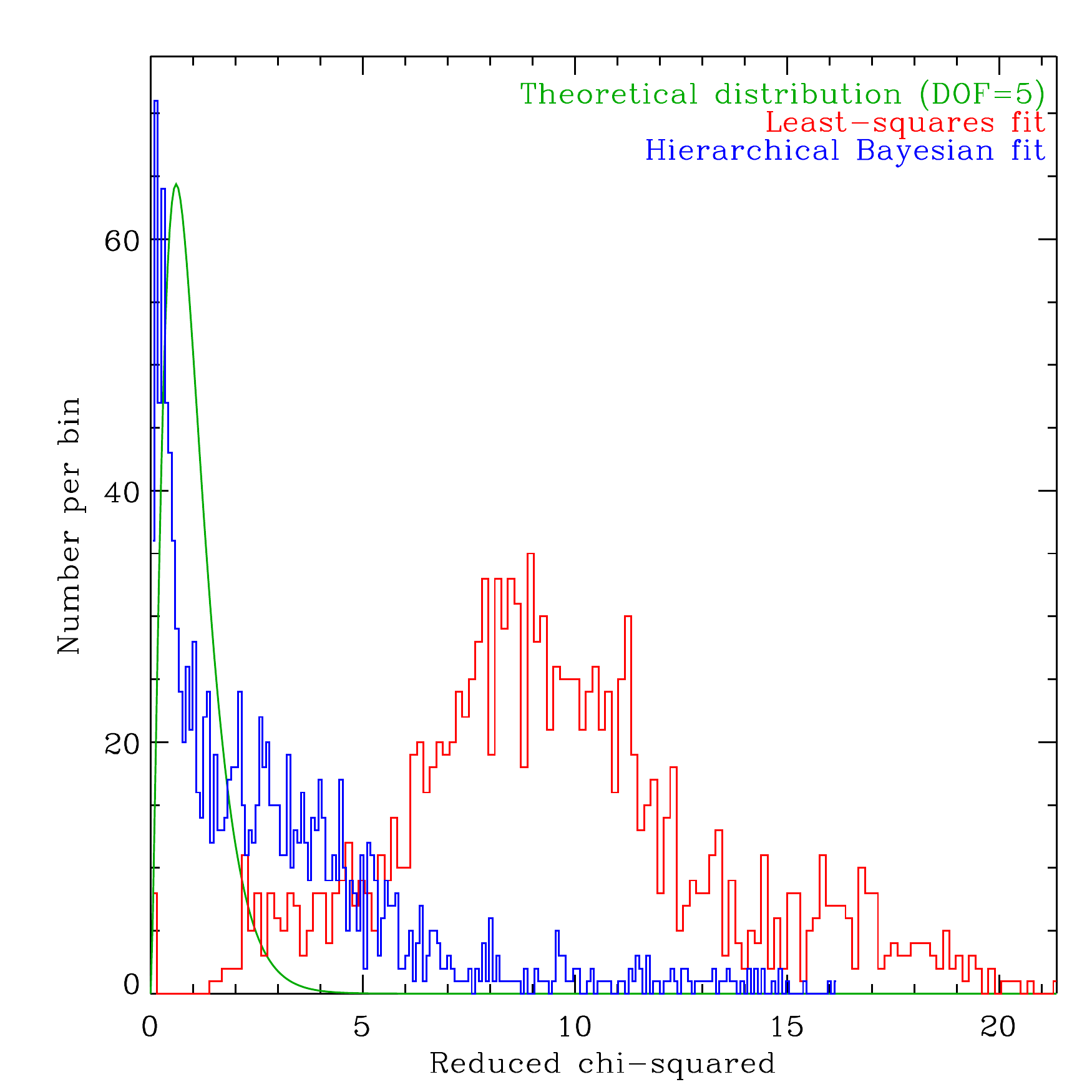}
        \end{center}
            \caption{Chi-squared distributions of all the pixels in used in the analysis, for LSM (red) and HB (blue), compared to the expected theoretical distribution (green).}
            \label{fig:dist_chi2red}
        \end{figure*}

        Indicated in Figs.~\ref{fig:LOri_SEDfit_Mdust}, \ref{fig:LOri_SEDfit_Mpah}, and \ref{fig:LOri_SEDfit_Mpahi}, the trends described in this chapter between dust mass and PAH mass, and the AME intensity were revealed through the HB analysis. 
        The increased scatter of best-fit points, produced by LSM, obscures the intrinsic relationships. With only LSM results, the dust SED fitting comparison would have been less clear. 
        Fig.~\ref{fig:dist_chi2red} highlights the difference between LSM in HB in terms of their chi-squared distributions.
         Such a comparison however requires us to note two caveats: chi-squared based analysis has been shown to be misleading in some cases \citep[see][]{andrae10}, and is not a rigorously interpretable parameter within the HB framework itself. 
         With these in mind, chi-squared distributions are still a useful way to investigate systematic differences between the two methods. 
         Fig.~\ref{fig:dist_chi2red} indicates that LSM tends to result in high chi-squared values.
         HB however produces values closer to the expected theoretical distribution, despite having a high chi-squared tail. 
         This tail is likely caused be remaining systematic residuals in the fit. 

\begin{figure*}
        \begin{center}
            \includegraphics[width=\textwidth]{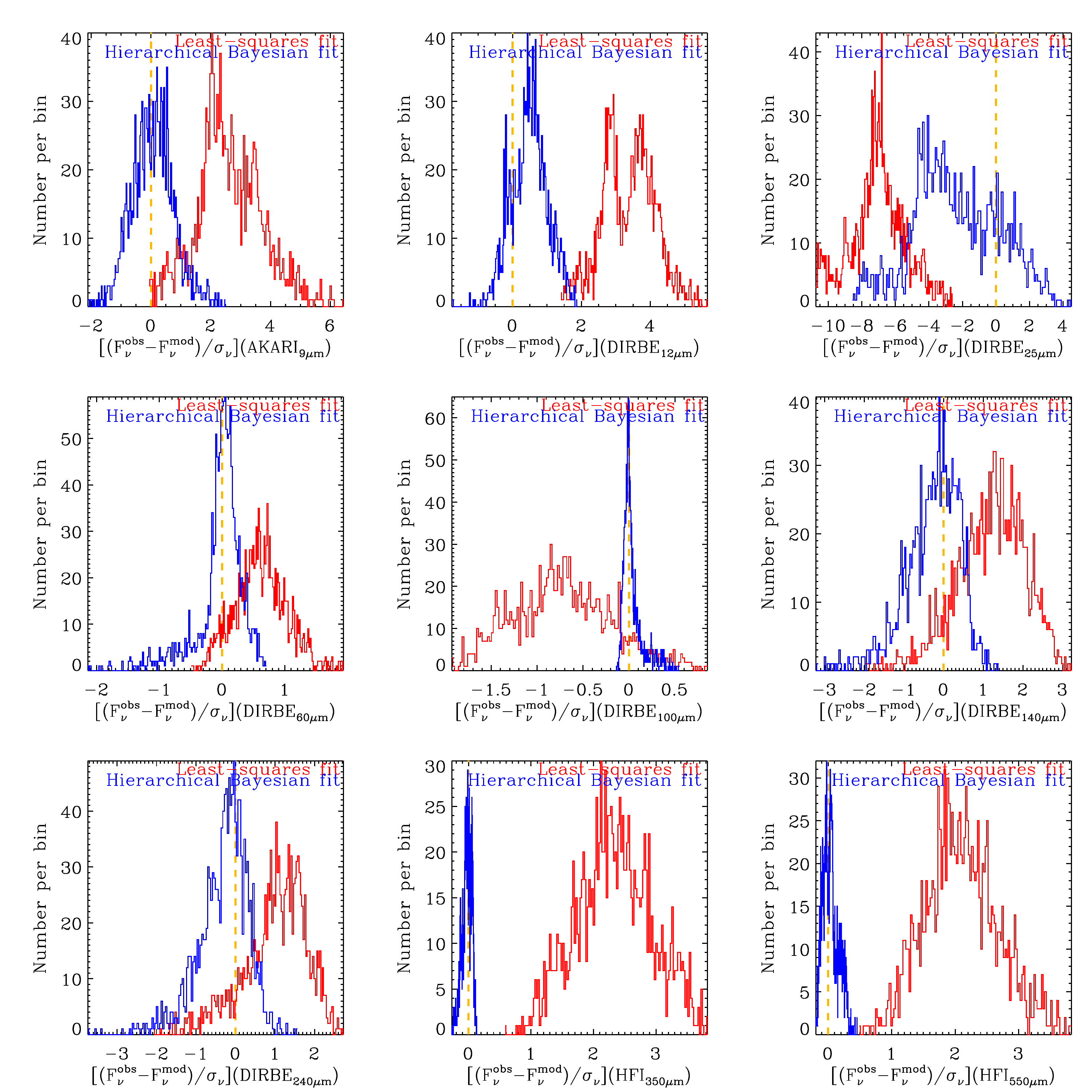}
        \end{center}
            \caption{Residual distributions per band when applying the LSM (red) vs. HB (blue) method, for all pixels in the analysis. Top row, from left to right: A9, D12, D25; Middle row: D60, D100, D140; Bottom row: D240, P857, P545. }
            \label{fig:residuals}
        \end{figure*}

        By studying the residuals of each band individually, one can investigate the origin of the high chi-squared tail of the HB distribution.
        Fig.~\ref{fig:residuals} compares the relative residuals of each band, for both LSM and HB.
        
        Fig.~\ref{fig:residuals} shows that HB leads to roughly symmetric residuals centered around 0, with almost no point beyond 3$\sigma$. The exception is the 25~$\mu$m band, which has a slightly skewed HB residual distribution. This indicates a model excess at this wavelength.

        In our model, the 25~$\mu$m band is dominated by the emission of small, out-of-equilibrium, amorphous carbon grains. 
        This grain population is distinct from the PAHs, dominating the 9 and 12~$\mu$m bands, and from the large silicate and carbon grains, dominating the FIR power. 
        Assuming there are no artifacts in the DIRBE data, these residuals are likely due to the fact that the size distribution we have assumed has an overabundant fraction of nanometric amorphous carbon grains. 
        These grains account for less than 10\% of the mass \citep[see][Appendix A.2]{galliano11}. 
        Thus, this residual is not likely to bias the parameters discussed in this paper. This applies to parameters constrained by FIR, $M_{\rm dust}$ and $U$; as well as those constrained by the 9 and 12~$\mu$m bands, $M_{\rm PAH}$, $M_{\rm PAH+}$.
        
        To our knowledge, this work is the first case of the HB approach being applied to a localized investigation of AME, and certainly the first time that the dust SED of $\lambda$~Orionis region has been investigated in such a way. 
        This raises questions about the future of LSM-based analysis in dust SED studies, and highlights the potential of this HB framework developed by \citet{galliano18a}.

      \subsection{PAH Ionization fraction}
          As described in Sect.\ref{sec:datasources}, relative variations between the A9 and I12 intensities could be explained, to some extent, by the fraction of PAHs that are charged, $f_{\rm PAH+}$. Spectroscopically, we cannot distinguish between PAH anions or cations. 
          However if spinning dust emission arises from anions, a better correlation with the mass of charged PAHs $M_{\rm PAH+}$ is expected. 
          However if the PAHs are positively charged, a stronger correlation with $M_{\rm PAH+}$ is not expected. 
          This is due to the rotational excitability of the PAHs: anions are more susceptible to rotational excitation by $H^{+}$ and $C^{+}$ collisions \citep{ali-haimoud10}.

          Examining $\lambda$~Orionis in intensity, we find that the A9 intensity correlates more strongly with AME than I12 or D12.  
          This is consistent with the spinning PAH hypothesis, and taken alone may indicate that the 6.2~$\mu$m feature emission from charged PAHs, may be a better predictor of AME intensity.

          As shown by the dust SED fitting however, the probability distributions (Fig.~\ref{fig:LOri_SEDfit_PDFs}) of $r_{\rm p, PAH+}$ do not indicate that ionized PAH mass correlates better with AME. 
          Attempts to estimate the $M_{\rm PAH+}$ based on the available data appear to only add noise relative to $r_{\rm p, PAH}$. 
          The means of the two distributions $r_{\rm p, PAH+}$  and $r_{\rm p, PAH}$ are similar and $r_{\rm p, PAH+}$ shows a wider distribution. 
          Thus the question of whether or not AME comes predominantly from charged PAHs remains open.

          The fact that A9 correlates more strongly than the 12~$\mu$m bands, at least suggests that this topic is worth further investigation. 
          What is clear from the MIR and AME morphology is a transition from a relatively PAH depleted, warmer, higher $U$ zone in the center and warm dust in the center of the HII region to a PAH-supporting zone as expected of a photo-dissociation region (PDR).

          \citet{andrews16} predict a transition of PAH species along such a radiation field gradient, from complete PAH destruction in harsh environments, to survival of (sufficiently large) PAH anions near the surface of molecular clouds. 
          Thus if our stronger correlation with A9 indicates charged PAHs, this could be consistent with PAH anions surviving in the portions of $\lambda$~Orionis which are emitting the strongest AME. 
          Future wide-area spectral mapping of the ${\lambda}$~Orionis region may be able to conclusively test for increased $M_{\rm PAH+}$ in regions with stronger AME. 
          This would also help us to understand the extent to which [NeII] emission may contribute to the I12 emission, and if this may lead to a relatively improved correlation between AME and A9. 
          Such studies would be strongly aided by higher resolution probing of spatial variations the AME spectral profile, such as those enabled by Q-U-I Joint Tenerife Experiment \citep{quijote12, santos15, santos17}.

          \section{Summary}
          In Sect.~\ref{sec:intro} we outlined the remaining ISM mystery that is AME-- its history of discovery, and attempts to this point to explain its source. 
          A general consensus was described, that the AME appears to come from interstellar dust grains, and a mechanism by which tiny rapidly rotating grains can produce AME was discussed as well as efforts by the community to confirm this mechanism. 
          All-sky works were reviewed and an argument for moving towards targeted studies of specific AME prominent regions was presented, specifically highlighting the $\lambda$~Orionis region as the target of this work. 
          In Sect.~\ref{sec:datasources} we reviewed the IR photometric data utilized for comparison to the AME data by the PC. 
          Specific benefits of employing the AKARI IRC 9~$\mu$m were highlighted, as this photometric bands provide uniquely complete coverage of PAH features, especially those from ionized PAHs which had not yet been explored with regards to AME.
          In Sect.~\ref{sec:dataproc} we described our data analysis strategy, which can be simply thought of as an attempt to put the IR and AME data on the same pixel grid, at the same angular resolution, for direct comparison across the $\lambda$~Orionis ring structure, and its central region. 
          In Sect.~\ref{sec:analysis} we introduced our analysis methods: an evolution of comparisons from simple IR to IR, and IR to AME cross-correlations, to a more robust IR-band to AME correlation test via bootstrap analysis to estimate statistical distributions of the correlation scores. 
          We then moved to a much more sophisticated analysis, via HB enabled dust SED model fitting and inference. 
          After describing the merits and justification for using HB-- mainly robust error propagation-- we demonstrated in Sect.~\ref{sec:results} how the AME intensities of pixels within $\lambda$~Orionis compare with dust SED properties derived from the model fitting. 
          To robustly infer which fitted dust properties gave weaker or stronger relationships, we calculated PDFs for each parameter of interest vs. the AME: 
          we were mostly interested in whether $M_{\rm PAH}$ or $M_{\rm dust}$ showed the stronger correlation with $I_{\rm AME}$. 
          Moreover we showed that in each comparison, normalizing $I_{\rm AME}$ by $U$ improves the correlations with dust parameters.
          In Sect.~\ref{sec:discussion} we argue based primarily on the HB dust SED fitting results that PAHs remain a likely carrier of the AME within $\lambda$~Orionis. 
          We also suggest that $U$ may be an implicit tracer of spinning dust excitation environments. 
          If we look simply at $I_{\rm AME}$ we may be ignoring important environmental factors of the AME-dust relationship.
          Finally we suggest that the specific role of ionized PAHs should be explored more carefully in the future, and that perhaps more emphasis should be placed on localized studies of AME rather than additional all-sky or large-scale studies.
          We stress that further work is needed to understand the relationship between particular morphologies of AME-prominent HII regions and level of spatial correlation between AME and PAHs.

\begin{ack}
  This research is based on observations with AKARI, a JAXA project with the participation of ESA. Additional research travel funding for this project were provided by the Japan Foundation for Promotion of Astronomy, the Astronomical Society of Japan, and the U. Tokyo Grad. School of Science Research Abroad Program (GRASP). A.C. Bell received living and educational supports for this project from the Japanese Ministry of Education, Culture, Sports, Scienc and Technology (MEXT). This work is supported in part by JSPS and CNRS under the Japan -- France Research Cooperative Program and JSPS KAKENHI Grant Numbers JP23244021 and JP18K03691.

  The authors would like to thank the following individuals for feedback on the manuscript and fruitful research discussions: Itsuki Sakon, John Livingston, Clive Dickinson, Bruce T. Draine, Brandon S. Hensley, Natalie Ysard, Olivier Bern\'e, Doug Marshall, Francois Boulanger, Anthony P. Jones, Steven J. Gibson, Amit Pathak, Mridusmita Buragohain, Ho-Gyu Lee and Mehdi Shibahara. The staff of the U.Tokyo International Liaison Office were helpful in acquiring research, education, and travel funding on behalf of A.C. Bell.

\end{ack}


\end{document}